%% file: TOI700.tex
\documentclass[twocolumn]{aastex63}
\usepackage{amsmath}
\usepackage{hyperref}
\usepackage{graphicx}
\usepackage{natbib}
\usepackage[scaled]{helvet}
\usepackage{url}
\bibpunct{(}{)}{;}{a}{}{,}
\usepackage{textcomp}
\usepackage{mathpazo}
\usepackage{xspace}
\usepackage{apjfonts}
\usepackage{mathrsfs}
\usepackage{verbatim}

\usepackage{color}
\usepackage[normalem]{ulem}

\newcommand{\bjdtdb}{\ensuremath{\rm {BJD_{TDB}}}}
\newcommand{\feh}{\ensuremath{\left[{\rm Fe}/{\rm H}\right]}}

\newcommand{\teff}{\ensuremath{T_{\rm eff}}\xspace}

\newcommand{\msun}{\ensuremath{\,M_\Sun}}
\newcommand{\rsun}{\ensuremath{\,R_\Sun}}
\newcommand{\lsun}{\ensuremath{\,L_\Sun}}

\newcommand{\rearth}{\ensuremath{\,R_{\rm \Earth}}\xspace}
\newcommand{\mearth}{\ensuremath{\,M_{\rm \Earth}}\xspace}
\newcommand{\re}{\ensuremath{\,R_{\rm \Earth}}\xspace}
\newcommand{\me}{\ensuremath{\,M_{\rm \Earth}}\xspace}
\newcommand{\fave}{\langle F \rangle}
\newcommand{\fluxcgs}{10$^9$ erg s$^{-1}$ cm$^{-2}$}

\newcommand{\Kepler}{{\it Kepler}}

\newcommand{\tess}{{\it TESS}}
\newcommand{\TESS}{{\it TESS}}

\newcommand{\Spitzer}{{\it Spitzer}}
\newcommand{\Gaia}{{\it Gaia}}

\newcommand{\ms}{\,m\,s$^{-1}$}
\newcommand{\cms}{\,cm\,s$^{-1}$}

\newcommand{\thisstar}{TOI-700\xspace}
\newcommand{\thistic}{TIC 150428135\xspace}
\newcommand{\mstar}{\ensuremath{M_{*}}}
\newcommand{\rstar}{\ensuremath{R_{*}}}

\newcommand{\radiusb}{$1.037^{+0.065}_{-0.064}$\xspace\re\xspace}
\newcommand{\radiusc}{$2.65^{+0.16}_{-0.15}$\xspace\re\xspace}
\newcommand{\radiusd}{$1.144^{+0.062}_{-0.061}$\xspace\re\xspace}

\newcommand{\be}{\begin{equation}}
\newcommand{\ee}{\end{equation}}

\newcommand{\starmass}{$0.415\pm0.020$}
\newcommand{\starradius}{$0.424\pm0.017$}

\begin{document}

%\title{\Spitzer\ Confirms the First habitable zone Earth-Sized Planet From \TESS}
\title{The First Habitable Zone Earth-Sized Planet From \emph{TESS} II: \emph{Spitzer} Confirms \thisstar\ d}

\input{affiliations.tex}
\input{authors.tex}

\shorttitle{A Habitable Zone Earth-Size Planet from \TESS}
\shortauthors{Rodriguez et al.}

\begin{abstract}
We present {\it Spitzer} 4.5$\mu$m observations of the transit of TOI-700 d, a habitable zone Earth-sized planet in a multiplanet system transiting a nearby M-dwarf star (TIC 150428135, 2MASS J06282325-6534456). TOI-700 d has a radius of $1.144^{+0.062}_{-0.061}R_\oplus$ and orbits within its host star's conservative habitable zone with a period of 37.42 days ($T_\mathrm{eq} \sim 269$~K). TOI-700 also hosts two small inner planets (R$_b$=$1.037^{+0.065}_{-0.064}R_\oplus$ \& R$_c$=$2.65^{+0.16}_{-0.15}R_\oplus$) with periods of 9.98 and 16.05 days, respectively. Our {\it Spitzer} observations confirm the {\it TESS} detection of TOI-700 d and remove any remaining doubt that it is a genuine planet. We analyze the {\it Spitzer} light curve combined with the 11 sectors of {\it TESS} observations and a transit of TOI-700 c from the LCOGT network to determine the full system parameters. Although studying the atmosphere of TOI-700 d is not likely feasible with upcoming facilities, it may be possible to measure the mass of TOI-700 d using state-of-the-art radial velocity instruments (expected RV semi-amplitude of $\sim$70~cm/s).
\end{abstract}

\section{Introduction}

Humans have wondered whether life exists elsewhere in the universe for centuries. Thanks to new technologies and rapid advancements in the study of exoplanets in the past few decades, we are making progress towards answering this question scientifically. So far, thousands of small exoplanets are known, most of which were discovered by the \Kepler\ mission \citep{Borucki:2010} and astronomers have taken the first steps towards probing their compositions \citep[e.g.][]{Dressing:2015b, Rogers:2015} and atmospheres \citep[e.g.][]{Kreidberg:2014}.  Now, more sensitive instruments \citep[e.g.][]{Szentgyorgyi:2016} and telescopes \citep[e.g.][]{Gardner:2006, Roberge:2018, Gaudi:2018} are planned with the eventual goal of detecting biosignatures, in an Earth-like planetary atmosphere.

%Using the \textit{Hubble Space Telescope} (HST) and \Spitzer\ Space Telescope, we have studied the atmospheres of giant planets and refined our observing and analysis techniques \citep{Madhusudhan:2019} to the point where we can study a select set of smaller planets \citep{Knutson:2014, Kreidberg:2019}. These early observations have revealed  We are starting to probe the patterns , we We have taken our first steps towards characterizing the compositions \citep{Dressing:2015b} and atmospheres \citep{Kreidberg:2014, benneke:2019, Tsiaras:2019} of small planets outside our own solar system. 

%We are planning to build more sensitive instruments \citep[e.g.][]{Szentgyorgyi:2016} and telescopes \citep[e.g.][]{Gardner:2006, Roberge:2018, Gaudi:2018} which will probe . with the eventual goal of detecting bio signatures, like H$^2$O and O$^2$, in a Earth-like planetary atmosphere.

Though the possibility of detecting biosignatures in the future seems real, the prospects remain uncertain. Statistical results from \Kepler\ have shown that small, habitable-zone planets are common around low-mass host stars \citep{Dressing:2015}, but it is not clear how much these planets resemble the Earth. \Kepler\ discovered potentially rocky habitable-zone planets around M-dwarf stars \citep{Quintana:2014}, but they orbit stars too faint for precise radial velocity measurements, so we do not know if they are rocky like the Earth or if they are shrouded by thick atmospheres, inhospitable to life as we know it. Our best constraints on which planets are rocky and which have thick envelopes come from observations of highly-irradiated, hot planets \citep{Rogers:2015, Wolfgang:2015}. So far, very few temperate planets that are similar to the size of the Earth orbit host stars bright enough to carry out precise mass measurements through radial velocities. % Even less is known about the atmospheres of terrestrial planets in temperate orbits. So far, no atmospheric features have been detected in the atmosphere of any rocky exoplanet \citep{Diamond-Lowe:2018}, and it is uncertain under which conditions these planets retain atmospheres at all \citep{Wordsworth:2015}. 

Learning more about small, temperate planets requires finding such planets around brighter stars. In April 2018, NASA's Transiting Exoplanet Survey Satellite (\tess) mission launched with precisely this goal. So far, \TESS\ has discovered over 1000 exoplanet candidates orbiting some of the brightest and closest stars to the Sun \citep[][N. Guerrero et al. \textit{in prep}]{Ricker:2015}. Most of these planets orbit close to their stars and have high equilibrium temperatures, which allow us to study planets in highly irradiated environments and probe atmospheric loss \citep[e.g.][]{Vanderspek:2019, Kreidberg:2019}. \tess\ is also expected to discover a small number of planets in temperate orbits around low-mass M-dwarf stars, possibly including rocky planets orbiting in their stars' circumstellar habitable zones \citep{Sullivan:2015, Kaltenegger:2019}. Though \TESS\ has detected slightly larger planets in temperate orbits \citep[e.g.][]{Gunther:2019} and terrestrial planets in hot orbits \citep[e.g.][]{Winters:2019}, so far no potentially rocky, habitable-zone planets have been reported. 

{In this paper, we confirm the first Earth-sized planet orbiting in its host star's habitable zone discovered by \TESS. The planet, \thisstar\ d, is only 22\% $\pm$ 7\% larger than Earth, orbits an M-dwarf star (0.415\msun, \thistic) located 31 parsecs from the Sun, and is a promising target for future observations to measure its mass. Unfortunately, atmospheric characterization will likely not be feasible with the James Webb Space Telescope (JWST, \citealp{Beichman:2014}). In a companion to this paper, \citet{Gilbert:2020} characterize the \thisstar\ system and statistically validate \thisstar\ d and two other planets in the system.  \thisstar\ d will likely be an attractive target for future observations, so independent confirmation of the planet's existence is valuable before investing large amounts of telescope resources. 
%including radial velocity observations to measure its mass. Since TOI-700 d, without a mass measurement it is best to have an independent confirmation of that the planet is real before investing large amounts of observing time. 
Here, we present \Spitzer\ Space Telescope observations that confirm \thisstar\ d is a transiting planet and help to refine our knowledge of its parameters.} Our paper is organized as follows: In \S2 we discuss the \tess\ and \Spitzer\ observations and reduction methods. We present our global analysis using EXOFASTv2 \citep{Eastman:2013, Eastman:2019} in \S3. We place \thisstar\ d in the context of presently known planets and examine future prospects for characterizing the planet's mass and atmosphere in \S4, and give our conclusions in \S5.

\begin{figure*}[ht] %  figure placement: here, top, bottom, or page
   \centering
   \includegraphics[width=1.0 \linewidth, angle =0, trim = 0 10.5in 0 0 ]{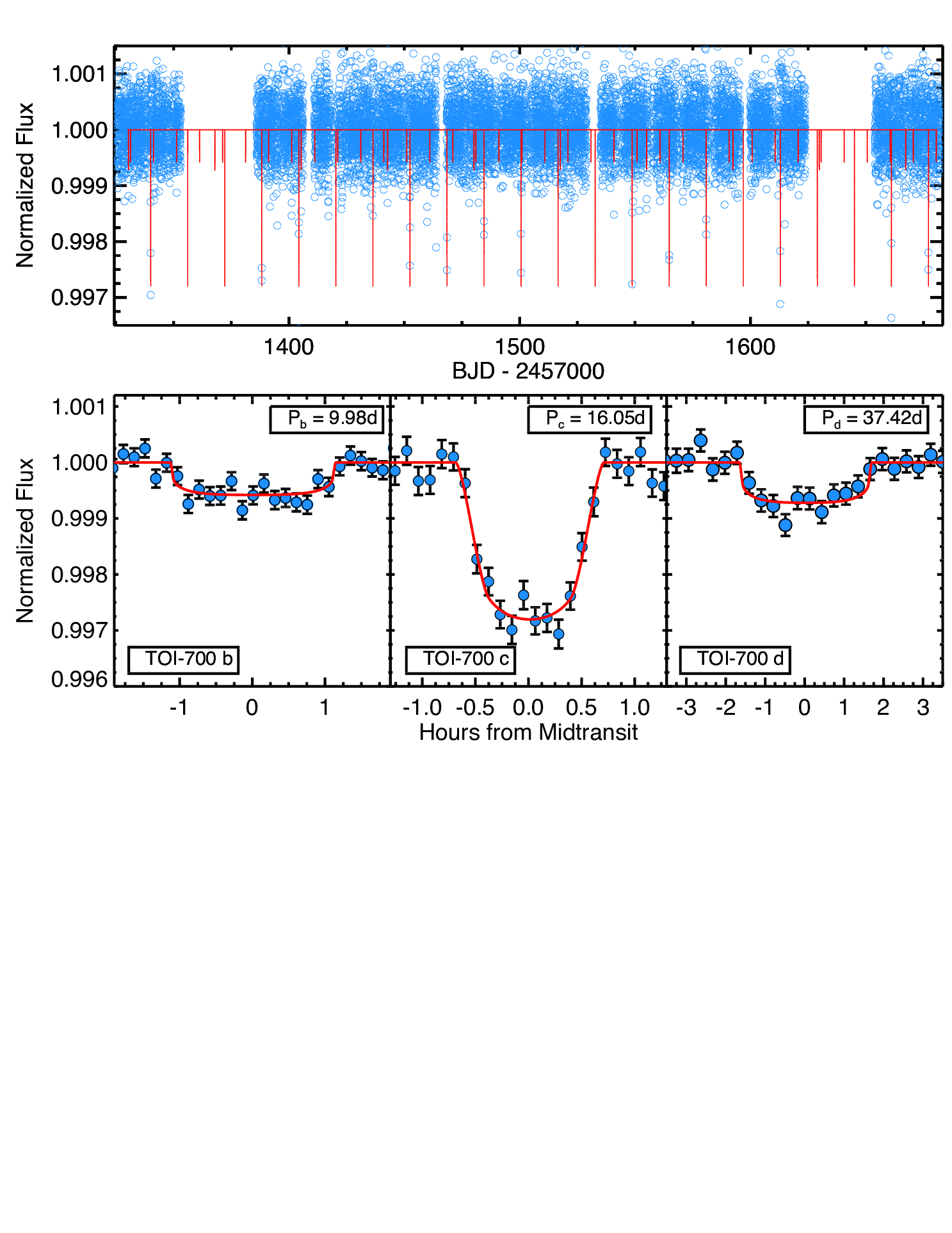} 
\caption{Photometric observations of the \thisstar\ system. The light curve has been processed to remove stellar variability and instrumental systematics. \textit{Top panel:} The blue open circles are the \tess\ observations binned to 30 minutes for all 11 sectors of \tess. Since \thisstar\ falls near \tess's continuous viewing zone, there are almost 10 months of data. \textit{Bottom panels:} The \tess\ transits for \textit{(left)} \thisstar\ b, \textit{(middle)} \thisstar\ c, and \textit{(right)} \thisstar\ d phase folded to the best fit ephemeris from the global fit (see \S\ref{sec:GlobalModel}). The \tess\ observations have been binned for visual clarity. The EXOFASTv2 transit model is shown in red in each panel.} 
   \label{fig:LC}
\end{figure*}

%The \tess\ and LCOGT transits for \textit{(top)} \thisstar\ b and \textit{(bottom)} \thisstar\ c. \textit{Bottom right panel:} The \textit{(top)} \tess\ and \textit{(bottom)} 4.5$\mu$m \Spitzer\ light curves (binned) phase folded to the best fit ephemeris from the global fit (see \S\ref{sec:GlobalModel}). The \tess\ and \Spitzer\ observations have been binned for visual clarity.

\begin{figure}[ht] %  figure placement: here, top, bottom, or page
   \centering
   \includegraphics[width=1.0 \linewidth, angle =0, trim = 0.3in 3.5in 4in 0 ]{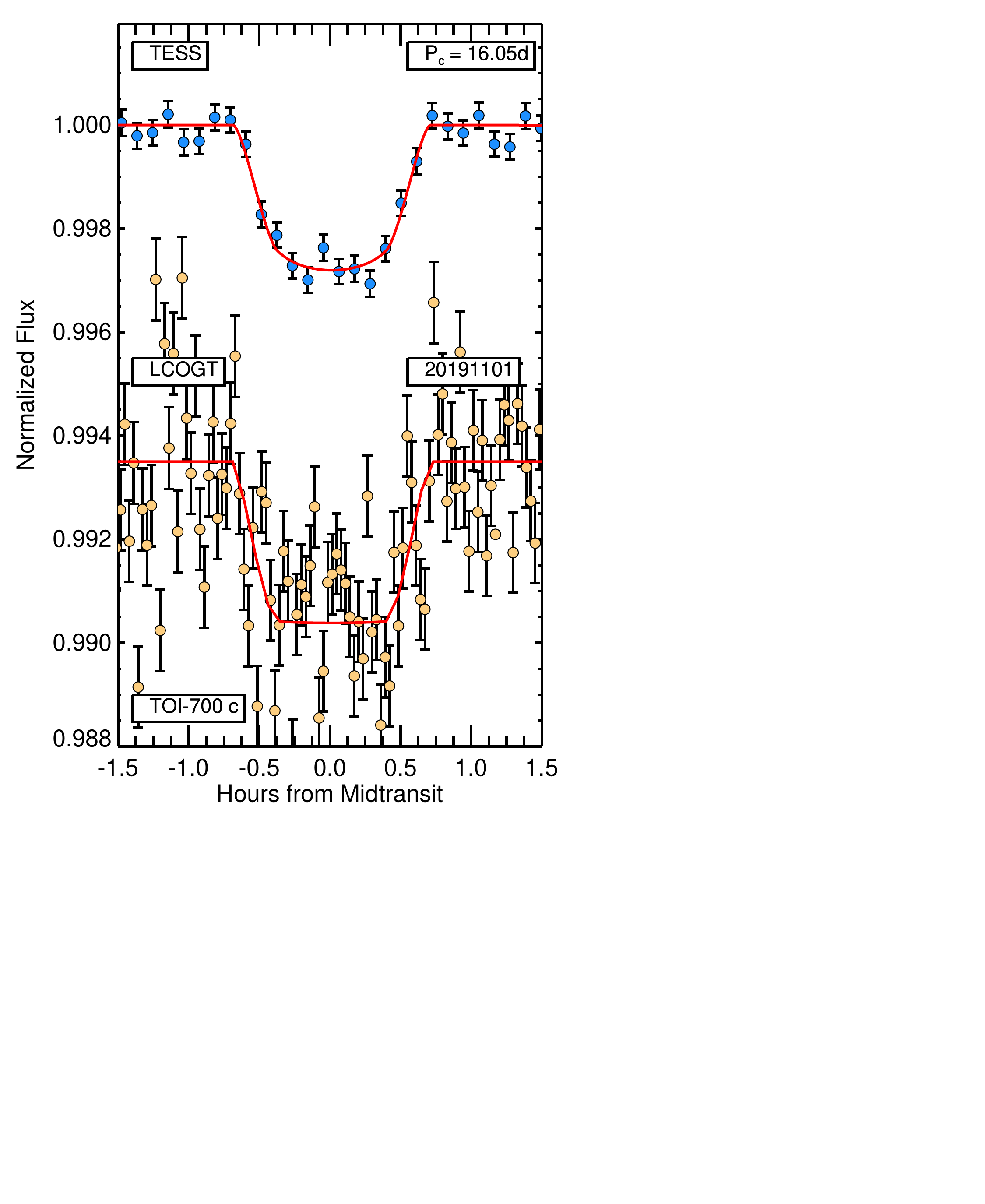} 
\caption{Photometric observations of the \thisstar\ c from \tess\ and LCO (UT 2019 November 01). The \tess\ and LCO observations have been binned for visual clarity. The EXOFASTv2 transit model for each data set is shown in red for each observation.}
   \label{fig:LCO}
\end{figure}

\begin{figure}[ht] %  figure placement: here, top, bottom, or page
   \centering
   \includegraphics[width=1.0 \linewidth, angle =0, trim = 0.3in 0.7in 4in 0in ]{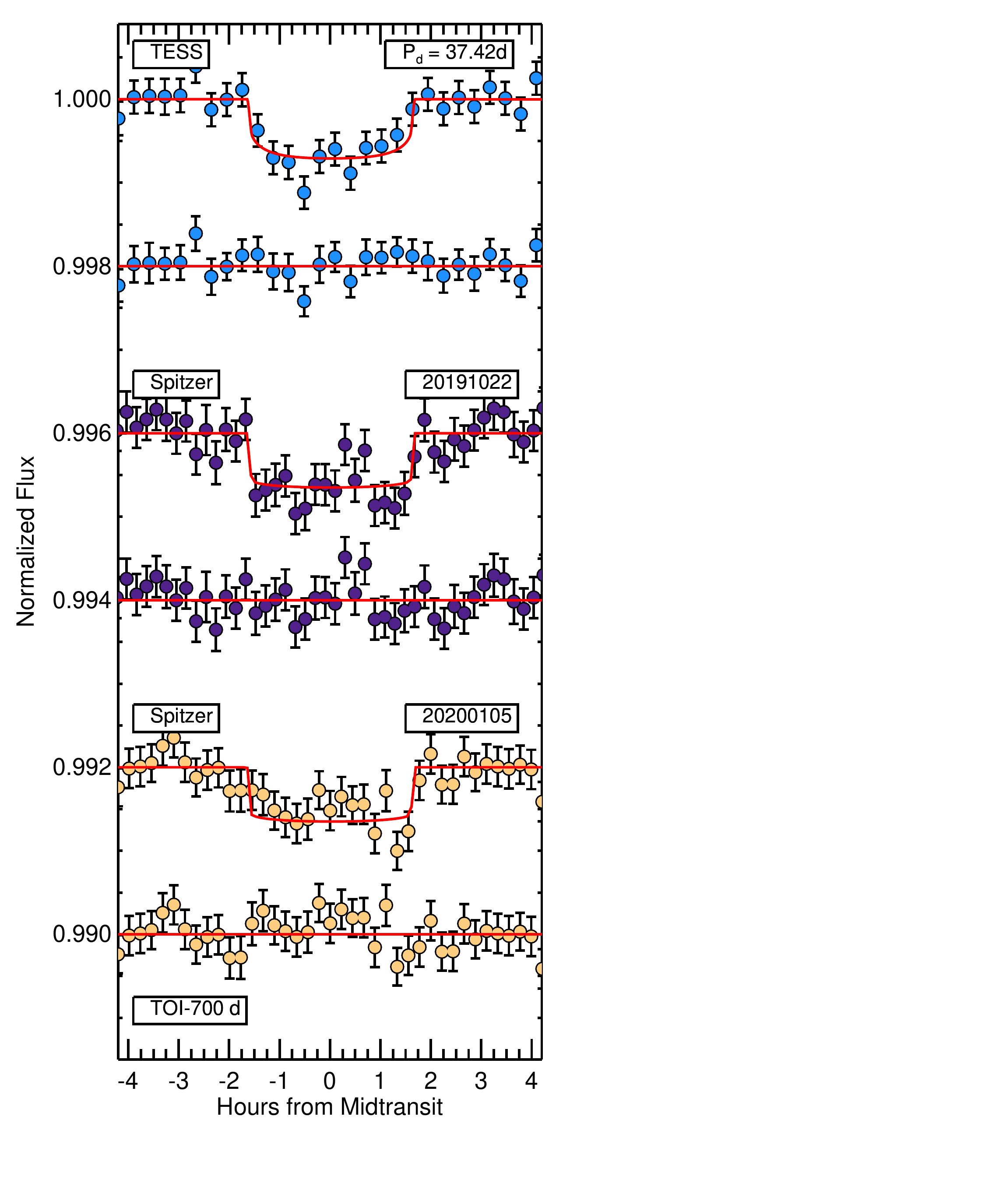} 
\caption{Photometric observations of the \thisstar\ d from \tess\ and Spitzer (UT 2019 October 22 \& UT 2020 January 05). The \tess\ and \Spitzer\ observations have been binned for visual clarity. The EXOFASTv2 transit model for each data set is shown in red for each observation. The residuals of our model fit are show below each lightcurve.} 
   \label{fig:spit}
\end{figure}

\section{Observations and Archival Data}
\label{Obs}
{In this section we present the observations used to confirm the small habitable-zone planet \thisstar\ d. In our analysis, we include the TESS discovery data and follow-up observations from \Spitzer\ and the Las Cumbres Observatory (LCO). In the companion to this paper, \citet{Gilbert:2020} characterized the \thisstar\ system and validated the planets using a wealth of additional follow-up observations including multiple spectroscopic observations, high-spatial resolution speckle imaging, and ground-based time series photometry. An additional companion paper, \citet{Suissa:2020}, explored plausible atmospheres for \thisstar\ d and the prospects for observing its atmosphere with future facilities. For brevity, we only describe the observations which directly feed into our EXOFASTv2 global analysis. See Table \ref{tbl:LitProps} for the literature magnitudes and kinematics. }

\input{lit_table.tex}

\subsection{{\it TESS} Photometry}
\label{sec:TESS}

\TESS\ observed \thisstar\ between 25 July 2018 and 17 July 2019. Because \thisstar\ is located near the southern ecliptic pole, it fell in a region of sky that was observed nearly continuously by \TESS. In total, \thisstar\ was observed during 11 \tess\ sectors. Though \thisstar\ is a nearby, bright dwarf star, it was not originally pre-selected for high-cadence \TESS\ observations because of incorrect catalog stellar parameters. It was, however, proposed as part of Guest Investigator proposal G011180 (PI Dressing), so pixel time series from a small region of the \tess\ CCD near \thisstar\ were downlinked with two-minute sampling. 

After the data were downlinked, they were reduced and analyzed by the Science Processing Operations Center (SPOC) pipeline, based at the NASA Ames Research Center \citep{Jenkins:2016}. The SPOC pipeline applies pixel-level calibrations to the data, identifies optimal photometric apertures, estimates flux contamination from other nearby stars, and extracts light curves. Instrumental artifacts are identified and removed from the light curves using the Presearch Data Conditioning (PDC) module (\citealt{Smith:2012}; \citealt{Stumpe:2014}). The processed light curves were searched for transits with the SPOC Transiting Planet Search (TPS, \citealt{Jenkins:2002}). 

Early searches of the \thisstar\ \tess\ light curves (both single sector and sectors 1-3) revealed some evidence for planetary transits, but these signals were not initially considered compelling enough to be promoted to the status of a planet candidate in the \TESS\ Object of Interest (TOI) catalog (N. Guerrero et al. \textit{in prep}). After data from \TESS\ sectors 1-6 were searched together, two planet candidates with 16.05 day and 37.42 day periods were detected and released in the online TOI catalog. A third planet candidate with period 9.98 days was detected in a subsequent search of data from \TESS\ sectors 1-9, and a final search of the \TESS\ full first-year dataset (11 sectors in total, see Figure \ref{fig:LC}) confirmed the detections \citep{Twicken:2018, Li:2019}.

These three candidates were investigated and characterized in detail by \citet{Gilbert:2020}, who statistically validated all three candidates as exoplanets. We note that it was more difficult to validate the planetary nature of the outer 37 day period planet candidate around \thisstar. The transit signal was detected by \TESS\ on only 8 occasions and was fairly weak with a Multiple Event Statistic (MES, a proxy for signal-to-noise ratio used by the SPOC pipeline) of 9.3. Experience from the \Kepler\ mission has shown that a significant fraction of planet candidates in the low signal-to-noise/few transit regime are false positives \citep{Mullally:2018, Burke:2019}. Although \citet{Gilbert:2020} showed that the false positive probability of \thisstar\ d (including instrumental false alarms) was below the threshold required for statistical validation, a small possibility remained that \thisstar\ d was spurious. Given the potential impact of the discovery, the likelihood that \thisstar\ d will be the target of future observations, and the fact that we are still working to fully understand systematic errors in \TESS\ data, we wanted an independent confirmation of the planet from another facility. We therefore proposed for observations with the \Spitzer\ Space Telescope to confirm the planet by detecting a transit at the predicted time.

\subsection{{\it LCOGT} Photometry of \thisstar\ c}
\label{sec:LCO}
As part of the \TESS\ Follow-up Observing Program (TFOP), a transit of \thisstar\ c was observed on UT 2019 November 01 in the z$^\prime$ band using one of the 1.0m Las Cumbres Observatory Global Telescope Network (LCOGT) telescopes located at the South African Astronomical Observatory (SAAO) \citep{Brown:2013}\footnote{https://lco.global}. The telescope is a Ritchey-Chretien Cassegrain with a 4k$\times$4k Sinistro CCD with a 27$\arcmin$ $\times$ 27$\arcmin$ field-of-view and a 0.39$\arcsec$ pixel scale. The observations were scheduled using the \tess\ Transit Finder that is built from the {\tt Tapir} software tool \citep{Jensen:2013}. The observations were reduced and the light curves were extracted using the {\tt AstroImageJ} software package \citep{Collins:2017}. The photometry was extracted using a 3.9$\arcsec$ aperture, as smaller apertures resulted in a significantly higher rms noise level. See Figure \ref{fig:LCO} for the plot of the LCOGT transit of \thisstar\ c.

\subsection{\Spitzer\ Photometry of \thisstar\ d}
\label{sec:spitzer}
Two follow up transits of \thisstar\ d were observed with the \Spitzer\ InfraRed Array Camera (IRAC, \citealt{Fazio:2004}) on UT 2019 October 22 and UT 2020 January 05 as part of a Director's Discretionary Time (DDT) proposal award (program number 14314, PI Vanderburg). Each observation was 8.9 hours in duration, with a two-second exposure time, and used Channel 2 on IRAC, which is equivalent to a photometric wavelength range of 4 -- 5 $\mu$m. Prior to the observations of \thisstar, a 30-minute burn-in sequence was conducted to allow the spacecraft to thermally equilibrate and the detector to asymptote to a steady state of charge trapping and release. For both the burn-in sequence and the time series observations, \thisstar\ was placed on the detector on a pixel that is known to have minimal sensitivity variations. 

We downloaded the \Spitzer\ observations from the archive, and reduced the Basic Calibrated Data (BCD, provided by the \Spitzer\ Science Center) using the custom aperture photometry routine developed by \citet{Cubillos:2013}. This analysis package (which is available open-source on GitHub\footnote{https://github.com/kevin218/POET}) fits a 2D Gaussian profile to the stellar image in each Spitzer exposure after upsampling by a factor of 5 in each spatial direction.  We identified and masked pixels with outlying values using an iterative sigma-clipping procedure and then summed the flux in each fixed aperture. We tested apertures with radii ranging from 2 to 4 pixels in 0.25 pixel steps, and found that a radius of 3.0 pixels minimizes the noise in each of the extracted light curves.  An annulus with an inner radius of 7 pixels and outer radius of 15 pixels was adopted for the determination of the median background value. 

%We tested aperture sizes from 2 to 4 pixels in 0.25-pixels steps for the target and found an optimal aperture of 3.0 pixels, which minimizes the noise in the extracted light curve. An annulus of 7 to 15 pixels was adopted to determine the median background flux value.

The dominating systematics for the 4.5 $\mu$m \Spitzer\ channel are intrapixel sensitivity variations \citep{Charbonneau:2005}. We therefore fitted for them by using the BiLinearly-Interpolated Subpixel Sensitivity (BLISS) map technique introduced by \citet{Stevenson:2012}. We describe the full \Spitzer\ light curve, $F(x,y,X,Y,t)$, by:
%The BLISS map technique assumes that the \Spitzer\ light curve, $F(x,y,X,Y,t)$, in its most general form can be described by:

%The complete flux model $F(x,y,X,Y,t)$ is therefore described by:

\begin{equation}
    F(x,y,X,Y,t) = F_s R(t) M(x,y) T(t) G(X,Y,t),
\end{equation}

\noindent where $F_s$ is the constant out-of-transit flux, $R(t)$ is the ramp model, $M(x,y)$ is the BLISS map with $(x,y)$ describing the position of the star on the detector, $T(t)$ is the \citet{Mandel:2002} transit model implemented in BATMAN \citep{Kreidberg:2015} and $G(X,Y,t)$ is a term fitting for variations in the pixel response function (PRF) using a 2D cubic with the gaussian widths $(X,Y)$. 

An initial fit with the BLISS map model revealed a clear transit in each of the \Spitzer\ light curves with the same depth and duration seen in the \TESS\ light curve. After detecting the transits, we adjusted our systematics correction to further optimize the \Spitzer\ light curve. The optimal resolution for BLISS mapping was found to be 0.01 pixels for the first observation and 0.008 pixels for the second. We also experimented with the complexity of the light curve model. In order to compare models with different numbers of free parameters, we used the Bayesian Information Criterion (BIC, \citealt{Schwarz:1978}, \citealt{Liddle:2007}). Combinations of a linear ramp $R(t)$ and PRF fits $G(X,Y,t)$ of different orders were tested. A significant increase in the BIC for those models showed that these more complex models are not justified. 

Our final model consisted only of the BLISS map, a constant and the BATMAN transit model. The latter has the following parameters: $T_0$, $R_P/R_*$, $P$, $a/R_*$, $\cos i$, $e$, $\omega_*$, $u_1$ and $u_2$ (see Table \ref{tbl:exofastv2} for a description of these parameters). As multiple transits were observed with \TESS\ and only two with \Spitzer, we fixed the period $P$ to the value determined by \TESS. 

Finally, we compared the \Spitzer\ model with a fit which fixes the system specific parameters ($P$, $a/R_*$, $\cos i$) to values from a fit of the \TESS\ observations only using EXOFASTv2. Both of these cases reproduce transit depths which are consistent with each other. The final fitted 4.5$\mu$m light curves from \Spitzer\ with the EXOFASTv2 global model are shown in Figure \ref{fig:spit}. \Spitzer\ independently detected the transit of \thisstar\ d with 12$\sigma$ confidence.

In order to account for the correlated noise in our light curves, we calculate $\beta$, using the ``time-averaging'' method \citep{Pont:2006, Winn:2007}. It scales the standard deviation of the data set with a factor $\beta$, which denotes the ratio between the actual achieved standard deviation of the binned residuals and the standard deviation in absence of red noise. By taking median values of this ratio for binnings between 15 and 30 minutes following \citet{Winn:2008}, we estimate $\beta_1 = 1.268$ for the first observation and $\beta_2 = 1.210$ for the second.

\input{EXOFAST_Fit_new.tex}

\section{EXOFAST\lowercase{v}2 Global Fits} 
\label{sec:GlobalModel}
To determine the full system parameters, and especially those of the habitable-zone Earth-sized planet, \thisstar\ d, we globally fit the photometric observations from eleven sectors of \tess, the observations from {\it Spitzer}, and a follow up transit of \thisstar\ c from the Las Cumbres Observatory using EXOFASTv2 \citep{Eastman:2013, Eastman:2019}. We removed low-frequency variability from the \TESS\ light curves by fitting the light curves with basis splines (with a 1.5 day knot spacing), ignoring points during the transits of the three planets, and iteratively excluding outliers \citep[see Fig. 3 from][]{Vanderburg:2014}. For computational efficiency, we averaged the two-second cadence \Spitzer\ light curve into one-minute bins. The \Spitzer\ light curve had been corrected for systematics as described in \S\ref{sec:spitzer}. 

EXOFASTv2 models planetary systems self consistently, so the transit parameters of each planet, \thisstar\ b, c, and d were fit simultaneously with, and informed by, their host star's parameters. Because \thisstar\ is a low-mass dwarf star, the stellar evolutionary and stellar atmospheric models embedded within EXOFASTv2 are not reliable. {Therefore, we use the absolute K-mag relations from \citet{Mann:2015, Mann:2019} to determine the mass and radius of \thisstar, and use these values with a conservative 5\% uncertainty as Gaussian priors on \rstar\ and \mstar\ of 0.419$\pm$0.021\rsun\ and 0.417$\pm$0.021 \msun. We note that these values are well within 1$\sigma$ of the mass and radius used for \thisstar\ from \citet{Gilbert:2020}.} We do not use the \citet{Claret:2017} limb-darkening tables within the global fit to constrain $u_{1}$ and $u_{2}$, and instead leave the limb-darkening parameters to be constrained by the transit light curves in each bandpass, as well as EXOFASTv2's built in uniform prior that only allows steps within the physical bounds identified by \citet{Kipping:2013} for any band: $u_1 + u_2 < 1$, $u_1 > 0$, and $u_1 + 2u_2 > 0$. We also place priors on \teff\ (3460$\pm$65 K) and \feh\ (-0.07$\pm$0.11 dex) from an analysis of spectroscopic observations using the Southern Astrophysical Research (SOAR) Telescope combined with an spectral energy distribution analysis (see \citet{Gilbert:2020}). Since priors are placed on \mstar\ and \rstar, the corresponding priors on \teff\ and \feh\ do not affect the fitted planet parameters, and are only used to derive quantities such as L$_{\star}$ and T$_{\rm eq}$, along with the predicted \Spitzer\ 3.6 and 4.5 $\mu$m eclipse depths shown in Table \ref{tbl:exofastv2}. To account for the possible correlated noise in the \Spitzer\ observations, we have modified the per point error for each \Spitzer\ light curve following \citet{Winn:2008}. Specifically,  the per point errors on for the \Spitzer\ light curves are derived by multiplying the $\beta$ parameter calculated for each \Spitzer\ light curve (see \S\ref{sec:spitzer}) and the light curve rms.  See Figures \ref{fig:LC}, \ref{fig:LCO}, and \ref{fig:spit} light curve plots of the \TESS, LCOGT, \Spitzer\ with the EXOFASTv2 models (light curves are available with this publication).

\begin{figure*}[ht!] %  figure placement: here, top, bottom, or page
   \centering
   \includegraphics[width=0.99\linewidth]{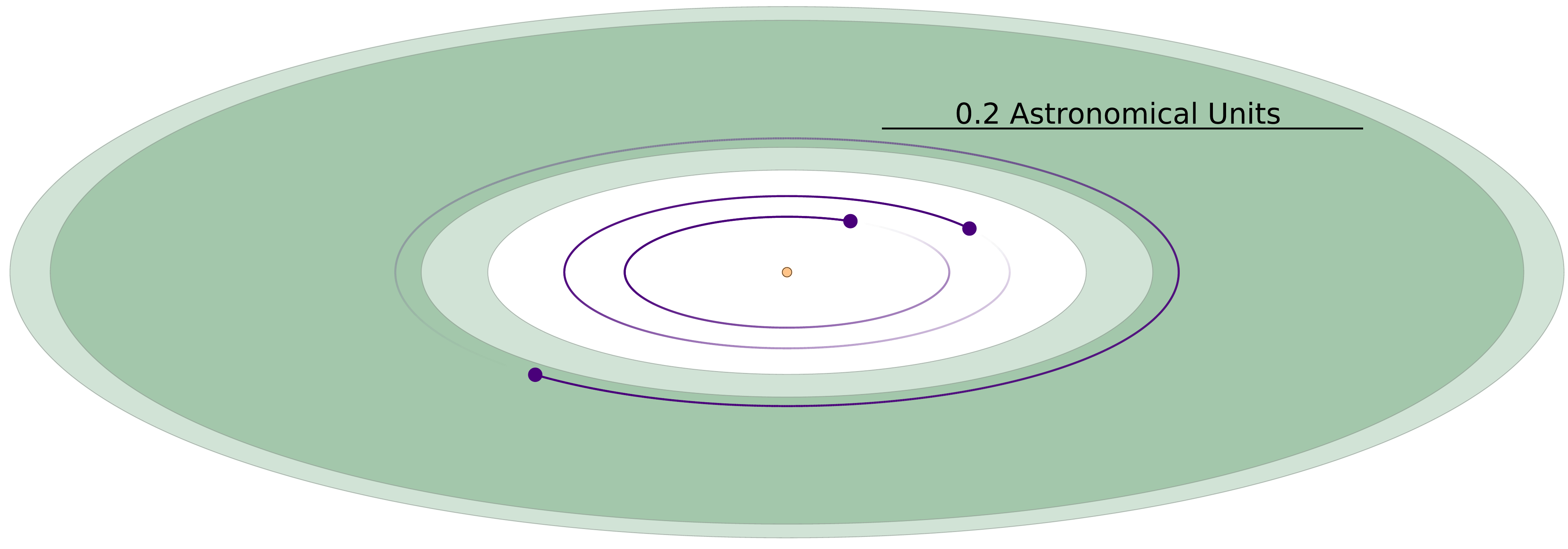} 
   \caption{Schematic of the \thisstar\ system, from the vantage of an observer inclined by 20 degrees from the plane of the system, showing the orbits of the planets (faded purple lines), the conservative habitable zone (dark green), and the optimistic extension to the habitable zone \citep[light green,][]{Kasting1993,Kane2012,Kopparapu2013,Kopparapu2014}. The size of the star \thisstar\ is shown to scale, but the sizes of the planets are not.} 
   \label{fig:hz}
\end{figure*}

Since we are simultaneously modeling photometric data for \thisstar\ d from different observatories, we performed some experiments to ensure that the data from \TESS\ and \Spitzer\ were consistent and neither significantly biased the transit depth. We do not perform these tests on the LCOGT observations because the LCOGT observations are far less precise. %Additionally, from our experience of jointly modeling \tess\ and ground-based observations, we have seen that the ground-based observations typically have a small influence on the results, other than improving the ephemeris as a result of an extended photometric baseline.
%EXOFASTv2 provides multiple ways to jointly model the \Spitzer\ and \tess\ photometry. The fit used in this paper jointly fits the  \Spitzer\ and \tess\ photometry to estimate a combined depth.
%To test whether or not our methodology was affecting the results or if we were detecting any significant difference in the depth of \thisstar\ d's transit, we jointly fit the {\it Spitzer} and \tess\ photometry allowing for dilution in either data set while using the other as the baseline for comparison.
We tested the \TESS\ and \Spitzer\  datasets for consistency by jointly fitting the two light curves, allowing for a dilution term in either data set while using the other as the baseline for comparison. Fitting for a dilution term in one bandpass effectively decoupled the measured transit depth from the two observatories. 
In all cases, the fit results were consistent with no significant additional dilution needed -- in other words, the transit depth measured by \TESS\ was consistent with that measured by \Spitzer. This is not surprising given that the SPOC pipeline accounts for blending from nearby stars in the \TESS\ data and that none of our follow up observations, including the \Spitzer\ images and high spatial resolution speckle observations, show evidence of any unknown additional stellar companions close enough to contaminate the \Spitzer\ photometry.

%Also, the SPOC pipeline to reduce the \tess\ observations (see \S\ref{sec:TESS}) accounts for all known companions and properly deblends the \tess\ light curve. 

%Can restore this when the tests finish
%We also ran two additional fits where we allowed EXOFASTv2 to fit the depths in each bandpass separately using its {\tt tdeltav} flag. We find a low-significance ($<$3$\sigma$) difference in the depths between \tess\ and \Spitzer, with \tess\ being a bit shallower. At the measured radius of planet d (1.21\rearth), it is not likely that \thisstar\ d has a large enough atmosphere to produce a detectable depth difference between the two filters, and we attribute any differences in the measured transit depths to red noise in the \TESS\ and \Spitzer\ light curves. In all cases, we get a planet radius for \thisstar\ d that is consistent at the 1$\sigma$ level with the results presented in Table \ref{tbl:exofastv2}.  

\section{Discussion}
\label{sec:discussion}
\thisstar\ is a compelling system for future characterization observations thanks to its relative proximity (d=31.075$\pm$0.038 pc) and brightness ($H=8.893\pm0.027$). All three planets orbiting \thisstar\ are sub-Neptune in size (R$_b$ = \radiusb\rearth, R$_c$ = \radiusc\rearth, and R$_d$ = \radiusd\rearth), and orbit with periods of 9.977, 16.0511, and 37.425 days. We note that the ratio of planet's orbital periods is 1.608 for c:b, and the d:c ratio is 2.33. These ratios are within 1\% of an 8:5 resonance c:b and 7:3 for d:c. See Section 6.3 of \citet{Gilbert:2020} which presents a detailed transit timing and photodynamical analysis. Their results suggest minimal perturbations on the orbit of planet d from planets b and c. We see no evidence of transit timing variations in the combined \Spitzer\ and \TESS\ global fit, supporting their result. We also see a slight mid-transit brightening during the Spitzer observations of \thisstar\ d. However, since it is similar in amplitude to other features in the light curve, not seen in the TESS observations, and that there is correlated noise in the Spitzer data, we do not believe it to be astrophysical in origin. We direct the reader to \citet{Gilbert:2020} for a detailed analysis and discussion of the \thisstar\ system as a whole (and the validation of all three planets in the system), and we focus the rest of our discussion on \thisstar\ d, the first habitable-zone Earth-sized planet from NASA's \tess\ mission.

\subsection{\Spitzer\ Confirmation of \thisstar\ d}
\label{confirmation}

While \citet{Gilbert:2020} were able to rule out most false positive scenarios for \thisstar\ d, a small probability remained that the planet was not real. Our \Spitzer\ observations rule out many of the remaining false positive scenarios for the planet candidate. Most importantly, we have now detected \thisstar\ d's transit with two different telescopes, so systematic errors in the \TESS\ light curve cannot be the source of the signal. Given \thisstar\ d's relatively low detection significance and small number of observed transits, instrumental artifacts were the greatest uncertainty in the validation of \thisstar\ d. The \Spitzer\ observations have retired this risk. 

%and improved our measurement of the transit shape
%We allowed different transit depths and limb darkening parameters for each bandpass, and placed no constraints on the host stellar properties.
\Spitzer\ also showed that the transit of \thisstar\ d is achromatic, which constrains blended companions and rejects additional astrophysical false positive scenarios. We measured the transit depths ($R_p^2/R_*^2$) in the \TESS\ and \Spitzer\ bandpasses by fitting the light curves simultaneously with \citet{Mandel:2002} models using an affine invariant Markov Chain Monte Carlo sampler \citep{GoodmanWeare2010}. We found that the \Spitzer\ transit (612 $\pm$ 44 ppm) is slightly shallower than the \TESS\ signal (677 $\pm$ 98 ppm). We constrained red contaminants by comparing the measured ratio of the \Spitzer/\TESS\ depths (0.90\ensuremath{^{+0.14}_{-0.12}}) to the expected ratios for a variety of cooler companions using Equation 5 from \citet{Desert:2015} and MIST isochrones \citep{Choi:2016}.  We rule out blends with very red, co-moving stars with mass less than 0.30 \msun\ (95\% confidence). The \Spitzer\ observations eliminate almost all remaining false positive scenarios for \thisstar\ d, especially instrumental artifacts. These data, combined with the observations and statistical validation presented by \citet{Gilbert:2020}, allow us to confidently pursue future observations.

\subsection{\Spitzer\ Improvement on \thisstar\ c and d's Parameters}
\label{sec:Spitimprov}
From our joint \tess\ and \Spitzer\ global fit, we are able to improve the properties of \thisstar\ d, relative to the results of an EXOFASTv2 fit that only included the \tess\ observations. This is similar to the fit that is presented in \citet{Gilbert:2020}, except they used the {\it exoplanet} software package \citep{Foreman-Mackey:2019}. This comparison removes any concern that any observed differences or improvements are caused by the analysis methodology. The \Spitzer\ observations took place on UT 2019 October 22 and UT 2020 January 05, 99 and 174 days after the end of the 11 \tess\ sectors in which \thisstar\ was observed (UT 2019 July 17), extending the total time baseline of observations by 49\%. The combined \Spitzer\ transits are  detected at $\sim14\sigma$ while the \TESS\ combined transits are detected at 7$\sigma$. This expansion in the photometric baseline yielded a 61\% improvement on the average precision on \thisstar\ d's orbital period. Additionally, thanks to its infrared capability (ideal for M-dwarfs like \thisstar) and larger aperture compared to \TESS, \Spitzer\ was able to reduce the uncertainty in $R_P/R_*$ for \thisstar\ d's by 39\%. Our analysis also included a transit of \thisstar\ c from LCOGT on UT 2019 November 01, 108 days after the end of the 11 sectors of \tess. Our analysis shows a 32\% improvement on the precision of \thisstar\ c's orbital period and a 9\% improvement on the planet's $R_P/R_*$.

The improved precision of these parameters will pay dividends as more follow-up observations of \thisstar\ d are conducted.  The improved orbital period measurement will help plan future transit observations of \thisstar\ d more efficiently, and our more precise radius measurement will be critical for understanding the planet's bulk composition. Decreasing radius uncertainties is particularly important for understanding the composition of rocky planets. For rocky planets (with some constant bulk iron/silicate abundance ratio), the planet's mass $m$ and radius $r$ are related by $m\propto r^{3.6}$ \citep{Zeng:2016}. This means that when inferring a rocky planet's iron/silicate ratio, a planet's radius must be known 3.6 times more precisely than its mass for the two observables to contribute equally. In other words, improving the radius precision will improve the ability of future radial velocity observations to determine \thisstar\ d's bulk composition.

\subsection{\thisstar's Habitable Zone}
\label{sec:hz}

The location of the habitable zone of planetary systems is calculated based on the premise that a planet similar to Earth could retain surface liquid water given sufficient atmospheric pressure \citep{Kasting1993,Kane2012,Kopparapu2013,Kopparapu2014}. Such calculations are sensitive to the precision of the stellar parameters \citep{Kane2014,Kane2018}; in particular the luminosity and effective temperature of the host star. Specifically, the habitable-zone boundaries are estimated using 1-dimensional cloud-free climate models that monitor the change in surface radiative balance for an Earth analog as a function of incident flux at infrared wavelengths. For this purpose, we utilize equations 4 and 5 along with the coefficients of table 1 from \citet{Kopparapu2014} to calculate the habitable-zone boundaries. The habitable zone is often described as either ``conservative'' with boundaries of runaway greenhouse and maximum greenhouse, or ``optimistic'' with boundaries determined from empirical assumptions of water prevalence on Venus and Mars \citep{Kane2016}. We use the stellar parameters shown in Table \ref{tbl:exofastv2} to calculate the extent of the \thisstar\ habitable zone for both the conservative and optimistic cases. Figure~\ref{fig:hz} shows a schematic of the \thisstar\ system comparing the orbits of the planets to the location of the conservative and optimistic habitable zones. \thisstar\ d's orbit lies confidently within the conservative habitable zone, and is small enough (only 20\% larger than the Earth) that it could be terrestrial \citep{Rogers:2015, Wolfgang:2015}. It is worth noting the caveat that there are various effects that influence the boundaries of the habitable zone. In the case of tidal locking, calculations from climate models have demonstrated that this generally has the effect of widening the habitable zone \citep{Yang2013,Yang2019}. Based on estimates of tidal locking time scales by \citet{Barnes2017}, the majority of habitable-zone terrestrial planets discovered by {\it TESS} are expected to be tidally locked for ages less than $\sim$1~Gyr, therefore increasing the confidence in the habitable-zone status of \thisstar\ d.

\subsection{Known habitable-zone Terrestrial-Sized Planets}
\label{sec:hab}
%With the confirmation of \thisstar\ d and it orbiting well within the conservative habitable zone of its host star, it now joins a very small population of habitable-zone terrestrial-sized planets. 
\thisstar\ d joins a very small population of presently-known habitable-zone terrestrial-sized planets. In this subsection, we compare \thisstar\ d to a sample of habitable-zone planets very similar in size to the Earth. Starting from a list of known small habitable-zone planets\footnote{\url{http://phl.upr.edu/projects/habitable-exoplanets-catalog}}, we identify planets smaller than 1.5\rearth, the radius below which hot planets orbiting M-dwarfs similar to \thisstar\ ($0.415\pm0.021$ \msun) tend to have rocky compositions \citep{Rogers:2015, Fulton:2017, Cloutier:2019}. After this cut, we are left with ten small habitable-zone\footnote{K2-72 e and TRAPPIST-1 d orbit in their star's optimistic habitable zones, while the others orbit within the conservative habitable zone.} planets: TRAPPIST-1 (d, e, f, \& g; \citealp{Gillon:2017}), \Kepler-186 f \citep{Quintana:2014}, \Kepler-1229 b \citep{Morton:2016}, \Kepler-442 b \citep{Torres:2015}, \Kepler-62 f \citep{Borucki:2013}, \Kepler-1649 c \citep{Vanderburg:2020} (this was recently announced and not in the known list at the time of this paper), and \thisstar\ d. We note that this cut removes all RV only planets and LHS-1140 b (1.73 \rearth), which has a density consistent with a rocky composition \citep{Dittmann:2017, Ment:2019}. Of the remaining planets, \thisstar\ d orbits the brightest host star by far. In the optical, \thisstar's \Gaia\ G-band magnitude (12.07) is 4.5 times brighter than the next brightest host (\Kepler-62, $G$ = 13.72), and in the near-infrared, \thisstar's K-band magnitude (8.63) is 4.6 times brighter than the next brightest host (TRAPPIST-1, $K$ = 10.30). \thisstar's apparent brightness makes it particularly attractive among small habitable-zone planets for follow-up observations. 

%K2-72 e \citep{Crossfield:2016, Dressing:2017},
%Therefore, our cut at 1.5 \rearth\ is focusing our comparison on system's very similar in size to the Earth. % moved this up to the beginning of the paragraph. 

%LHS-1140 b (1.73 \rearth), which is the only planet in this regime that has a mass measurement, and its density is -- that's not true, the trappist planets have mass measurements too

%Of the remaining systems, only two host stars have an H-band brightness below $H<$10 are \thisstar\ and TRAPPIST-1 \citep{Gillon:2017}.% I don't think this is true: I find TRAPPIST-1's H-mag = 10.718 . 

%We emphasize the $H$-band magnitude since any future characterization observations would mostly be in the Near-IR, using the Hubble Space Telescope's Wide Field Camera 3 instrument and/or the upcoming suite of instruments that will be available for exoplanet observations on the James Webb Space Telescope \citep{Beichman:2014, Stevenson:2016}. %Since atmospheres are out of the question for HST, may as well not mention this

% which is consistent with a rocky composition when looking mass-radius distribution of terrestrial planets with host stars similar to

\subsection{Future Radial Velocity Observations of \thisstar\ d}
\label{sec:RVaccess}

\begin{figure*}[ht!] %  figure placement: here, top, bottom, or page
   \centering
   \includegraphics[width=0.99\linewidth]{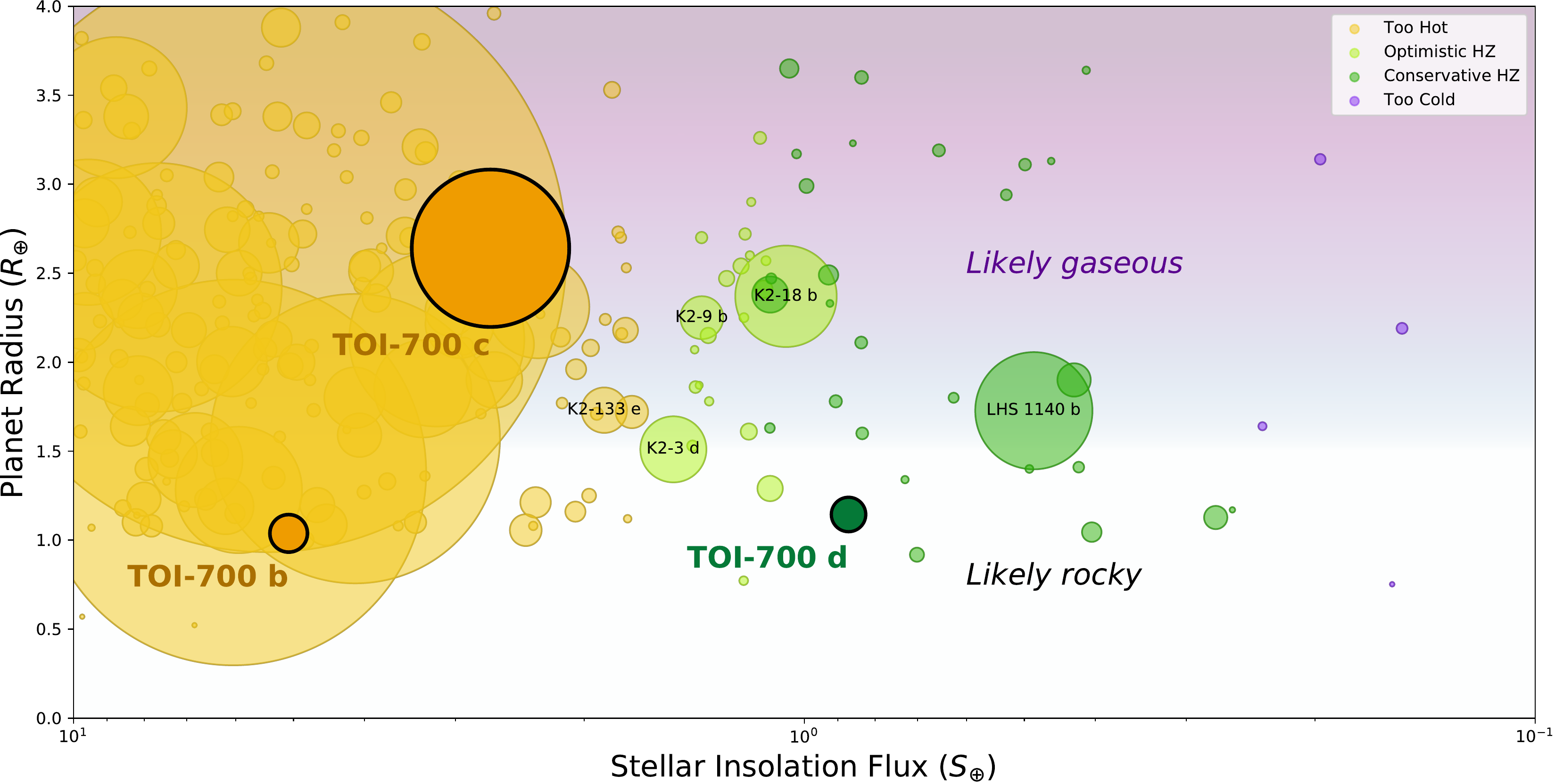} 
   \caption{\thisstar\ in the context of known exoplanets. Using data from the NASA Exoplanet Archive \citep{Akeson:2013}, we plot the radius of known exoplanets versus the incident flux they receive from their host stars. The symbols are color coded based on their position relative to the circumstellar habitable zone, where yellow dots represent planets too hot to have liquid water, purple dots represent planets too cold to have liquid water, and light and dark green points represent planets in the optimistic and conservative habitable zones, respectively. The background shading indicates whether a planet is likely rocky or gaseous based on its size alone, following \citep{Rogers:2015, Wolfgang:2015}. We used the polynomial expressions from \citet{Kopparapu2014} to determine the boundaries of the optimistic and conservative habitable zones for each host star. The area of the symbol is inversely proportional to the amount of observing time required to measure the planet's mass with radial velocity observations -- larger points are easier measurements (see \S\ref{sec:RVaccess}). In terms of RV accessibility, \thisstar\ d stands out as one of the best habitable-zone planets with a radii small enough that they are likely rocky. We note that LHS 1140 b is more accessible based on our metrics and has a measured mass consistent with it being rocky \citep{Dittmann:2017, Ment:2019}.} 
   \label{fig:RVaccess}
\end{figure*}

Among small, habitable-zone planets, \thisstar\ d is well-suited for precise radial velocity observations to measure its mass and confirm/rule out a rocky composition. Its host star is a quiet M-dwarf (0.415\msun) with no large photometric variations observed in the full \tess\ light curve. The host star appears to be relatively inactive with a long rotation period of 54$\pm$0.8 days. With no signs of activity in the spectra and a rotation period that differs from the orbital period, \thisstar\ should be well suited for RV follow up (we direct the reader to \citet{Gilbert:2020} for a thorough discussion on the stellar classification, including spectral analysis). Using the \citet{Chen:2017} mass/radius relation\footnote{If we adopt the \citet{Weiss:2014} mass-radius relation, we get similar masses for each planet: M$_b$ = 1.1\mearth, M$_c$ = 6.7\mearth, and M$_d$ = 2.2\mearth.}, our EXOFASTv2 model reports mass estimates for \thisstar\ b, c, and d to be M$_b$ = $1.25^{+1.00}_{-0.35}$\mearth, M$_c$ = $7.9^{+2.7}_{-1.8}$\mearth, and M$_d$ = $1.94^{+0.69}_{-0.57}$\mearth, which correspond to radial velocity semi-amplitudes of $0.67^{+0.54}_{-0.19}$, $3.61^{+1.3}_{-0.82}$, and $0.68^{+0.24}_{-0.20}$ \ms. Figure \ref{fig:RVaccess} compares the radial velocity accessibility of \thisstar\ d to other known transiting exoplanets. The symbol size is inversely proportional to a simple metric estimating the amount of observing time required to detect each planet's RV signal (assuming the \citealt{Weiss:2014} mass/radius relation):

\begin{equation}
Point\ Size \propto K^2 \times\ 10^{-0.4 G}\\
\end{equation}

\noindent Where K is the semi-amplitude and G is the \Gaia\ $G$-band magnitude of the host star. By this metric, \thisstar\ d is the best small habitable-zone (conservative) planet (R$_p$ $<$ 1.5 \re) for RV observations. Detecting the RV signal of \thisstar\ d will be challenging, but is within the current capabilities of the most precise spectrographs like the Echelle SPectrograph for Rocky Exoplanets and Stable Spectroscopic Observations (ESPRESSO) on the VLT \citep{Pepe:2010}. ESPRESSO is stable enough to detect the planet's $\approx 80$ \cms\ signal and \thisstar\ is bright enough that ESPRESSO should achieve $\approx$ 70 \cms\ photon-limited precision in 1-hour exposures  \citep[][and private communication]{Pepe:2014, Faria:2019}.

% (see Figure \ref{sec:discussion})

\subsection{Future Atmospheric Characterization of \thisstar\ d}
\label{sec:JWST}
Despite its favorable properties, \thisstar\ d will be a challenging target for transit spectroscopy observations to search for biosignatures or other molecules in its atmosphere in the near future. To assess the feasibility of detecting these features, we simulated JWST spectra for the planet (assuming Earth-like and CO$_2$ dominated atmospheres) with Pandexo \citep{Batalha:2017} using the NIRSpec/G235M observing mode. This mode provides the highest signal-to-noise for the transmission spectra of rocky planets around M-dwarfs \citep{Morley:2017}. Assuming photon noise limited observations, distinguishing such features from a featureless spectrum at 5$\sigma$ confidence would require data spanning more than 200 transits ( $\gtrsim 1000$ hours), equivalent to observing every single transit of \thisstar\ d for the first 20 years after JWST's launch. It would also require an order of magnitude higher precision on the transit depth measurements than has ever been achieved \citep{Line:2016}. \thisstar\ d requires significantly more observing time than the TRAPPIST-1 planets because of the relatively small planet-to-star radius ratio. We direct the reader to paper III in this series that does a much more in depth analysis of \thisstar\ d's possible atmosphere, including a 3-D general circulation model of plausible atmospheres and their detectability using future observatories \citep{Suissa:2020}. While follow-up JWST observations are not practical, the discovery of this planet motivates the development of future large-aperture observing facilities capable of sub-10 ppm measurement precision in the near-infrared.

% corresponding to over 1000 hours of JWST time

%Caroline and Laura

%\begin{figure*}
%\includegraphics[width=6.5in]{toi700systemsmall.pdf}
%\caption{A top-down view of the \thisstar\ system, showing the orbits of the planets (solid lines), the conservative HZ (light green), and the optimistic extension to the HZ (dark green). The scale of the figure is 0.42~AU along each side.}
%\label{fig:hz}
%\end{figure*}
  %\caption{A top-down view of the \thisstar\ system, showing the orbits of the planets (solid lines), the conservative HZ (light green), and the optimistic extension to the HZ (dark green). The scale of the figure is 0.42~AU along each side.}

\section{Conclusion}
\label{sec:conclusion}
We present new \Spitzer\ observations confirming the planetary nature of the \thisstar\ d, a habitable-zone Earth-sized planet located within a multiplanet system. \thisstar\ is a early M-dwarf (M$_{\star}$ = \starmass\ \msun\ and R$_{\star}$ = \starradius\ \rsun) located 31.1 pc from the Sun. {Using a combination of high spatial resolution speckle imaging, spectroscopic observations from CHIRON, and ground-based seeing limited photometry from the TESS Followup Observing Program (TFOP), \citep{Gilbert:2020} were able to statistically validate the planetary nature of \thisstar\ b, c, and d. Although the calculated false positive probability was  low enough for the planet to be statistically validated, we sought independent transit confirmation with \Spitzer\ given the importance of the discovery and the relatively low S/N of the transit signal from \TESS. Our 4.5$\mu$m \Spitzer\ observations conclusively confirmed the transit of \thisstar\ d, ruling out any remaining instrumental origin for the signal and solidifying its validation.}

We model the \TESS\ and \Spitzer\ photometry to determine the full system parameters.  \thisstar\ hosts two Earth-sized planets and a sub-Neptune (R$_b$ = $1.044^{+0.065}_{-0.063}$ \rearth, R$_c$ =$2.64^{+0.16}_{-0.14}$ \rearth, and R$_d$ =$1.220^{+0.073}_{-0.063}$ \rearth) with periods of P$_b$ = $9.97702^{+0.00024}_{-0.00028}$, P$_c$ =$16.051108^{+0.000062}_{-0.000064}$, and P$_d$ = $37.42469^{+0.00033}_{-0.00042}$ days. \thisstar\ d is located well within the conservative habitable zone for its host star, and is the first habitable-zone Earth-sized planet discovered from NASA's \tess\ mission. \thisstar\ is the brightest known host of a transiting habitable-zone Earth-sized planet discovered to date. 

Although atmospheric characterization is likely out of reach of current and upcoming facilities, \thisstar\ d provides a rare opportunity to measure the planet's mass with state of the art facilities like ESPRESSO on the VLT \citep{Pepe:2010}. Future observations should focus on measuring the mass of all three planets to gain insight into whether or not Earth-sized planets around low-mass stars are similar to the Earth.  The \TESS\ mission was recently selected for its first extended mission, which will begin in the summer of 2020. According to the draft observing schedule, \thisstar\ will be observed in 11 more sectors during \tess\ cycle 3\footnote{\url{https://heasarc.gsfc.nasa.gov/cgi-bin/tess/webtess/wtv.py}} providing a great opportunity to refine the ephemerides and parameters of the three known planets, and possibly detect additional planets in the system that would enhance our understanding of \thisstar's architecture.

\acknowledgements

AV's work was performed under contract with the California Institute of Technology (Caltech)/Jet Propulsion Laboratory (JPL) funded by NASA through the Sagan Fellowship Program executed by the NASA Exoplanet Science Institute. MNG acknowledges support from MIT's Kavli Institute as a Juan Carlos Torres Fellow. CDD acknowledges support from the NASA TESS Guest Investigator Program through Grant 80NSSC18K1583. EDL is thankful for support from GSFC Sellers Exoplanet Environments Collaboration (SEEC), which is funded by the NASA Planetary Science Divisions Internal Scientist Funding Model. JNW thanks the Heising-Simons Foundation for support. EAG thanks the LSSTC Data Science Fellowship Program, which is funded by LSSTC, NSF Cybertraining Grant \#1829740, the Brinson Foundation, and the Moore Foundation; her participation in the program has benefited this work. EAG is thankful for support from GSFC Sellers Exoplanet Environments Collaboration (SEEC), which is funded by the NASA Planetary Science Division's Internal Scientist Funding Model.

This work is based on observations made with the \Spitzer\ Space Telescope, which is operated by the Jet Propulsion Laboratory, California Institute of Technology under a contract with NASA. Support for this work was provided by NASA through an award issued by JPL/Caltech. This research has made use of SAO/NASA's Astrophysics Data System Bibliographic Services. This research has made use of the SIMBAD database, operated at CDS, Strasbourg, France. This work has made use of data from the European Space Agency (ESA) mission {\it Gaia} (\url{https://www.cosmos.esa.int/gaia}), processed by the {\it Gaia} Data Processing and Analysis Consortium (DPAC, \url{https://www.cosmos.esa.int/web/gaia/dpac/consortium}). Funding for the DPAC has been provided by national institutions, in particular the institutions participating in the {\it Gaia} Multilateral Agreement. This work makes use of observations from the Las Cumbres Observatory Global Telescope Network. We thank Kevin Stevenson for making the POET pipeline open-source and freely available on GitHub.

Funding for the {\it TESS} mission is provided by NASA's Science Mission directorate. We acknowledge the use of public {\it TESS} Alert data from pipelines at the {\it TESS} Science Office and at the {\it TESS} Science Processing Operations Center. This research has made use of the NASA Exoplanet Archive and the Exoplanet Follow-up Observation Program website, which are operated by the California Institute of Technology, under contract with the National Aeronautics and Space Administration under the Exoplanet Exploration Program. This paper includes data collected by the {\it TESS} mission, which are publicly available from the Mikulski Archive for Space Telescopes (MAST). This paper includes observations obtained under Gemini program GN-2018B-LP-101. Resources supporting this work were provided by the NASA High-End Computing (HEC) Program through the NASA Advanced Supercomputing (NAS) Division at Ames Research Center for the production of the SPOC data products.

\software{EXOFASTv2 \citep{Eastman:2013, Eastman:2019}, AstroImageJ \citep{Collins:2017}, BATMAN \citep{Kreidberg:2015}, Forecaster \citep{Chen:2017}}
\facilities{TESS, Spitzer, LCOGT,Gaia, MAST}

\bibliographystyle{apj}

\bibliography{refs}

%\input{ep248435473.best_as_of_04-02.tex}
%%%%%%%%%%%%%%%%%%%%%%%%%%%%%%%%%%%%%%%%%%%%%%%%%%%%%%%%%%%%%%%%%%%%%%%%%

% \appendix
% \section{Analytic Modeling of TTVs}
% \label{sec:appendix:ttv}

\end{document}

%% file: affiliations.tex
\newcommand{\cfa}{Center for Astrophysics \textbar \ Harvard \& Smithsonian, 60 Garden St, Cambridge, MA 02138, USA}
\newcommand{\umich}{Astronomy Department, University of Michigan, 1085 S University Avenue, Ann Arbor, MI 48109, USA}
\newcommand{\utaustin}{Department of Astronomy, The University of Texas at Austin, Austin, TX 78712, USA}
\newcommand{\MIT}{Department of Physics and Kavli Institute for Astrophysics and Space Research, Massachusetts Institute of Technology, Cambridge, MA 02139, USA}
\newcommand{\MITEPS}{Department of Earth, Atmospheric and Planetary Sciences, Massachusetts Institute of Technology,  Cambridge,  MA 02139, USA}
\newcommand{\uflorida}{Department of Astronomy, University of Florida, 211 Bryant Space Science Center, Gainesville, FL, 32611, USA}
\newcommand{\riverside}{Department of Earth and Planetary Sciences, University of California, Riverside, CA 92521, USA}
\newcommand{\usq}{University of Southern Queensland, West St, Darling Heights QLD 4350, Australia}
\newcommand{\ames}{NASA Ames Research Center, Moffett Field, CA, 94035, USA}
\newcommand{\geneva}{Observatoire de l’Universit\'e de Gen\`eve, 51 chemin des Maillettes,
1290 Versoix, Switzerland}
\newcommand{\uw}{Astronomy Department, University of Washington, Seattle, WA 98195 USA}
\newcommand{\warwick}{Deptartment of Physics, University of Warwick, Gibbet Hill Road, Coventry CV4 7AL, UK}
\newcommand{\warwickceh}{Centre for Exoplanets and Habitability, University of Warwick, Gibbet Hill Road, Coventry CV4 7AL, UK}
\newcommand{\princeton}{Department of Astrophysical Sciences, Princeton University, 4 Ivy Lane, Princeton, NJ, 08544, USA}
\newcommand{\liege}{Space Sciences, Technologies and Astrophysics Research (STAR) Institute, Universit\'e de Li\`ege, 19C All\'ee du 6 Ao\^ut, 4000 Li\`ege, Belgium}
\newcommand{\vanderbilt}{Department of Physics and Astronomy, Vanderbilt University, Nashville, TN 37235, USA}
\newcommand{\fisk}{Department of Physics, Fisk University, 1000 17th Avenue North, Nashville, TN 37208, USA}
\newcommand{\columbia}{Department of Astronomy, Columbia University, 550 West 120th Street, New York, NY 10027, USA}
\newcommand{\toronto}{Dunlap Institute for Astronomy and Astrophysics, University of Toronto, Ontario M5S 3H4, Canada}
\newcommand{\unc}{Department of Physics and Astronomy, University of North Carolina at Chapel Hill, Chapel Hill, NC 27599, USA}
\newcommand{\iac}{Instituto de Astrof\'isica de Canarias (IAC), E-38205 La Laguna, Tenerife, Spain}
\newcommand{\lalaguna}{Departamento de Astrof\'isica, Universidad de La Laguna (ULL), E-38206 La Laguna, Tenerife, Spain}
\newcommand{\louisville}{Department of Physics and Astronomy, University of Louisville, Louisville, KY 40292, USA}
\newcommand{\aavso}{American Association of Variable Star Observers, 49 Bay State Road, Cambridge, MA 02138, USA}
\newcommand{\utokyo}{The University of Tokyo, 7-3-1 Hongo, Bunky\={o}, Tokyo 113-8654, Japan}
\newcommand{\naoj}{National Astronomical Observatory of Japan, 2-21-1 Osawa, Mitaka, Tokyo 181-8588, Japan}
\newcommand{\jstpresto}{JST, PRESTO, 7-3-1 Hongo, Bunkyo-ku, Tokyo 113-0033, Japan}
\newcommand{\astrobiojapan}{Astrobiology Center, 2-21-1 Osawa, Mitaka, Tokyo 181-8588, Japan}
\newcommand{\ctio}{Cerro Tololo Inter-American Observatory, Casilla 603, La Serena, Chile}
\newcommand{\nexsci}{Caltech/IPAC, 1200 E. California Blvd. Pasadena, CA 91125, USA}
\newcommand{\ucsc}{Department of Astronomy and Astrophysics, University of
California, Santa Cruz, CA 95064, USA}
\newcommand{\gsfc}{Exoplanets and Stellar Astrophysics Laboratory, Code 667, NASA Goddard Space Flight Center, Greenbelt, MD 20771, USA}
\newcommand{\usno}{U.S. Naval Observatory, Washington, DC 20392, USA}
\newcommand{\gsfcsellers}{GSFC Sellers Exoplanet Environments Collaboration, NASA Goddard Space Flight Center, Greenbelt, MD 20771 }

\newcommand{\sgtinc}{SGT, Inc./NASA AMES Research Center, Mailstop 269-3, Bldg T35C, P.O. Box 1, Moffett Field, CA 94035, USA}
\newcommand{\chile}{Center of Astro-Engineering UC, Pontificia Universidad Cat\'olica de Chile, Av. Vicu\~{n}a Mackenna 4860, 7820436 Macul, Santiago, Chile}
\newcommand{\Pontificia}{Instituto de Astrof\'isica, Pontificia Universidad Cat\'olica de Chile, Av.\ Vicu\~na Mackenna 4860, Macul, Santiago, Chile}
\newcommand{\Millennium}{Millennium Institute for Astrophysics, Chile}
\newcommand{\maxplank}{Max-Planck-Institut f\"ur Astronomie, K\"onigstuhl 17, Heidelberg 69117, Germany}
\newcommand{\utdallas}{Department of Physics, The University of Texas at Dallas, 800 West
Campbell Road, Richardson, TX 75080-3021 USA}
\newcommand{\MauryLewin}{Maury Lewin Astronomical Observatory, Glendora, CA 91741, USA}
\newcommand{\umbc}{University of Maryland, Baltimore County, 1000 Hilltop Circle, Baltimore, MD 21250, USA}
\newcommand{\osu}{Department of Astronomy, The Ohio State University, 140 West 18th Avenue, Columbus, OH 43210, USA}
\newcommand{\MITAA}{Department of Aeronautics and Astronautics, MIT, 77 Massachusetts Avenue, Cambridge, MA 02139, USA}
\newcommand{\openu}{School of Physical Sciences, The Open University, Milton Keynes MK7 6AA, UK}
\newcommand{\swarthmore}{Department of Physics and Astronomy, Swarthmore College, Swarthmore, PA 19081, USA}
\newcommand{\seti}{SETI Institute, Mountain View, CA 94043, USA}
\newcommand{\lehigh}{Department of Physics, Lehigh University, 16 Memorial Drive East, Bethlehem, PA 18015, USA}
\newcommand{\utah}{Department of Physics and Astronomy, University of Utah, 115 South 1400 East, Salt Lake City, UT 84112, USA}
\newcommand{\USNA}{Department of Physics, United States Naval Academy, 572C Holloway Rd., Annapolis, MD 21402, USA}
\newcommand{\DTM}{Department of Terrestrial Magnetism, Carnegie Institution for Science, 5241 Broad Branch Road, NW, Washington, DC 20015, USA}
\newcommand{\UPenn}{The University of Pennsylvania, Department of Physics and Astronomy, Philadelphia, PA, 19104, USA}
\newcommand{\montana}{Department of Physics and Astronomy, University of Montana, 32 Campus Drive, No. 1080, Missoula, MT 59812 USA}
\newcommand{\psu}{Department of Astronomy \& Astrophysics, The Pennsylvania State University, 525 Davey Lab, University Park, PA 16802, USA}
\newcommand{\psust}{Center for Exoplanets and Habitable Worlds, The Pennsylvania State University, 525 Davey Lab, University Park, PA 16802, USA}
\newcommand{\Kutztown}{Department of Physical Sciences, Kutztown University, Kutztown, PA 19530, USA}
\newcommand{\udel}{Department of Physics \& Astronomy, University of Delaware, Newark, DE 19716, USA}
\newcommand{\Westminster}{Department of Physics, Westminster College, New Wilmington, PA 16172}
\newcommand{\steward}{Department of Astronomy and Steward Observatory, University of Arizona, Tucson, AZ 85721, USA}
\newcommand{\saao}{South African Astronomical Observatory, PO Box 9, Observatory, 7935, Cape Town, South Africa}
\newcommand{\salt}{Southern African Large Telescope, PO Box 9, Observatory, 7935, Cape Town, South Africa}
\newcommand{\ssl}{Societ\`{a} Astronomica Lunae, Italy}
\newcommand{\spot}{Spot Observatory, Nashville, TN 37206, USA}
\newcommand{\txamGP}{George P.\ and Cynthia Woods Mitchell Institute for Fundamental Physics and Astronomy, Texas A\&M University, College Station, TX77843 USA}
\newcommand{\txam}{Department of Physics and Astronomy, Texas A\&M university, College Station, TX 77843 USA}
\newcommand{\wellesley}{Department of Astronomy, Wellesley College, Wellesley, MA 02481, USA}
\newcommand{\byu}{Department of Physics and Astronomy, Brigham Young University, Provo, UT 84602, USA}
\newcommand{\Hazelwood}{Hazelwood Observatory, Churchill, Victoria, Australia}
\newcommand{\pest}{Perth Exoplanet Survey Telescope}
\newcommand{\Winer}{Winer Observatory, PO Box 797, Sonoita, AZ 85637, USA}
\newcommand{\icpo}{Ivan Curtis Private Observatory}
\newcommand{\elsauce}{El Sauce Observatory, Chile}
\newcommand{\crow}{Atalaia Group \& CROW Observatory, Portalegre, Portugal}
\newcommand{\dfus}{Dipartimento di Fisica ``E.R.Caianiello'', Universit\`a di Salerno, Via Giovanni Paolo II 132, Fisciano 84084, Italy}
\newcommand{\indfn}{Istituto Nazionale di Fisica Nucleare, Napoli, Italy}
\newcommand{\sotes}{Gabriel Murawski Private Observatory (SOTES)}
\newcommand{\chicago}{Department of Astronomy and Astrophysics, University of Chicago, Chicago, IL 60637, USA}
\newcommand{\dartmouth}{Department of Physics and Astronomy, Dartmouth College, Hanover, NH, USA}
\newcommand{\Tsinghua}{Department of Astronomy and Tsinghua Centre for Astrophysics, Tsinghua University, Beijing 100084, China}
\newcommand{\uibk}{Universit\"at Innsbruck, Institut f\"ur Astro- und Teilchenphysik, Technikerstra{\ss}e 25, 6020 Innsbruck, Austria}
\newcommand{\bu}{Department of Astronomy \& Institute for Astrophysical Research, Boston University, 725 Commonwealth Avenue, Boston, MA 02215, USA}
\newcommand{\colorado}{Department of Astrophysical and Planetary Sciences, University of Colorado, Boulder, CO 80309, USA}
\newcommand{\noqsi}{Noqsi Aerospace, Ltd., 15 Blanchard Avenue, Billerica, MA 01821, USA}
\newcommand{\berkely}{Department of Astronomy, University of California Berkeley, Berkeley, CA 94720-3411, USA}
\newcommand{\mcgill}{Department of Physics and McGill Space Institute, McGill University, 3550 rue University, Montreal, QC, H3A 2T8, Canada
}
\newcommand{\alder}{The Adler Planetarium, 1300 South Lakeshore Drive, Chicago, IL 60605, USA}
%Insert these before the institution affiliation command
\newcommand{\torres}{\altaffiliation{Juan Carlos Torres Fellow}}
\newcommand{\sagan}{\altaffiliation{NASA Sagan Fellow}}
\newcommand{\bernoulli}{\altaffiliation{Bernoulli fellow}}
\newcommand{\gruber}{\altaffiliation{Gruber fellow}}
\newcommand{\kavli}{\altaffiliation{Kavli Fellow}}
\newcommand{\peg}{\altaffiliation{51 Pegasi b Fellow}}
\newcommand{\pappalardo}{\altaffiliation{Pappalardo Fellow}}
\newcommand{\hubble}{\altaffiliation{NASA Hubble Fellow}}

%% file: authors.tex
\correspondingauthor{Joseph E. Rodriguez} 
\email{joseph.rodriguez@cfa.harvard.edu}

\author[0000-0001-8812-0565]{Joseph E. Rodriguez}
\affiliation{\cfa}

\author[0000-0001-7246-5438]{Andrew Vanderburg}%Agreed
\sagan
\affiliation{\utaustin}

\author[0000-0003-0562-6750]{Sebastian Zieba} %Agreed
\affiliation{\uibk}

\author[0000-0003-0514-1147]{Laura Kreidberg} %Agreed
\affiliation{\cfa}

\author[0000-0002-4404-0456.]{Caroline V. Morley} %agreed
\affiliation{\utaustin}

\author[0000-0003-3773-5142]{Jason D. Eastman} %Agreed
\affiliation{\cfa}

\author[0000-0002-7084-0529]{Stephen R. Kane}%Agreed
\affiliation{\riverside}

\author{Alton Spencer} %agreed
\affiliation{Danbury High School, Danbury, CT 06811, USA}

\author[0000-0002-8964-8377]{Samuel N. Quinn}%Agreed
\affiliation{\cfa}

\author[0000-0001-5383-9393]{Ryan Cloutier} %Agreed
\affiliation{\cfa}

\author[0000-0003-0918-7484]{Chelsea X. Huang} %agreed
\torres
\affiliation{\MIT}

\author[0000-0001-6588-9574]{Karen A. Collins} %Agreed
\affiliation{\cfa}

\author[0000-0003-3654-1602]{Andrew W. Mann} %Agreed
\affiliation{\unc}

\author[0000-0002-0388-8004]{Emily Gilbert} %Agreed
\affiliation{\chicago}
\affiliation{\alder}
\affiliation{\gsfc}
\affiliation{\gsfcsellers}

\author{Joshua E. Schlieder} %Agreed
\affiliation{\gsfc}
\affiliation{\gsfcsellers}

\author[0000-0003-1309-2904]{Elisa V. Quintana} %Agreed
\affiliation{\gsfc}
\affiliation{\gsfcsellers}

\author[0000-0001-7139-2724]{Thomas Barclay} %Agreed
\affiliation{\gsfc}
\affiliation{\umbc}
\affiliation{\gsfcsellers}

\author[0000-0003-4471-1042]{Gabrielle Suissa}%Agreed
\affiliation{\gsfc}
\affiliation{\gsfcsellers}
\affiliation{Goddard Earth Sciences Technology and Research (GESTAR), Universities Space Research Association, Columbia, Maryland, USA}

\author[0000-0002-5893-2471]{Ravi kumar Kopparapu}%Agreed
\affiliation{\gsfc}
\affiliation{\gsfcsellers}
\affiliation{NASA NExSS Virtual Planetary Laboratory, Box 951580, Seattle, WA 98195}

\author[0000-0001-8189-0233]{Courtney D. Dressing}%Agreed
\affiliation{\berkely}

% Author list Alphabetical here 

\author{George R. Ricker} %Agreed
\affiliation{\MIT}

\author{Roland K. Vanderspek} %Agreed
\affiliation{\MIT}

\author[0000-0001-9911-7388]{David W. Latham} %agreed
\affiliation{\cfa}

\author[0000-0002-6892-6948]{Sara Seager} %Agreed
\affiliation{\MIT}
\affiliation{\MITEPS}
\affiliation{\MITAA}

\author[0000-0002-4265-047X]{Joshua N. Winn} %agreed
\affiliation{\princeton}

\author[0000-0002-4715-9460]{Jon M. Jenkins} %agreed
\affiliation{\ames}

\author[0000-0002-3321-4924]{Zachory Berta-Thompson}%Agreed %agreed
\affiliation{\colorado}

\author{Patricia T. Boyd} %Agreed
\affiliation{\gsfc}
\affiliation{\gsfcsellers}

\author[0000-0002-9003-484X]{David Charbonneau} %agreed
\affiliation{\cfa}

\author{Douglas A. Caldwell}%Agreed
\affiliation{\seti}
\affiliation{\ames}

\author[0000-0002-6246-2310]{Eugene Chiang}%Agreed
\affiliation{\berkely}

\author[0000-0002-8035-4778]{Jessie L. Christiansen}%agreed
\affiliation{\nexsci}

\author[0000-0002-5741-3047]{David R. Ciardi}  %Agreed
\affiliation{\nexsci}

\author[0000-0001-8020-7121]{Knicole D.\ Col\'on} %agreed
\affiliation{\gsfc}
\affiliation{\gsfcsellers}

%\author[0000-0001-8274-6639]{Hannah Diamond-Lowe} %declined
%\affiliation{\cfa}

\author[0000-0003-2996-8421]{John Doty} %agreed
\affiliation{\noqsi}

\author[0000-0002-4503-9705]{Tianjun Gan} %agreed
\affiliation{\Tsinghua}

\author[0000-0002-5169-9427]{Natalia Guerrero} %Agreed
\affiliation{\MIT}

\author[0000-0002-3164-9086]{Maximilian~N.~G{\"u}nther}  %agreed
\torres
\affiliation{\MIT}

\author[0000-0002-1228-9820]{Eve J. Lee}%Agreed
\affiliation{\mcgill}

\author[0000-0001-8172-0453]{Alan M. Levine} %agreed
\affiliation{\MIT}

\author[0000-0002-7727-4603]{Eric Lopez}%Agreed
\affiliation{\gsfc}
\affiliation{\gsfcsellers}

\author[0000-0002-0638-8822]{Philip S. Muirhead} %agreed
\affiliation{\bu}

\author{Elisabeth Newton} %agreed
\affiliation{\dartmouth}

\author[0000-0003-4724-745X]{Mark E. Rose}%Agreed
\affiliation{\ames}

%\author[0000-0002-4829-7101]{Pamela~Rowden} %Agreed
%\affiliation{\openu}

\author[0000-0002-6778-7552]{Joseph D. Twicken} %SPOC
\affiliation{\seti}
\affiliation{\ames}

\author{Jesus Noel Villase\~{n}or} %agreed
\affiliation{\MIT}

%\author[0000-0002-4891-3517]{George Zhou} 
%\hubble
%\affiliation{\cfa}

%% file: lit_table.tex
\begin{table}
\scriptsize
\setlength{\tabcolsep}{2pt}
\centering
\caption{Properties of \thisstar}
\begin{tabular}{llcc}
  \hline
  \hline
Other identifiers\dotfill & \\
\multicolumn{3}{c}{TIC 150428135} \\
\multicolumn{3}{c}{2MASS J06282325-6534456} \\
\multicolumn{3}{c}{WISE J062823.05-653443.7}\\
\hline
\hline
Parameter & Description & Value & Source\\
\hline 
$\alpha_{J2000}$\dotfill	&Right Ascension (RA)\dotfill & 06:28:23.22878& 1	\\
$\delta_{J2000}$\dotfill	&Declination (Dec)\dotfill & 	-65:34:45.52157& 1	\\
\\
$l$\dotfill     & Galactic Longitude\dotfill & 275.4682095$^\circ$ & 1\\
$b$\dotfill     & Galactic Latitude\dotfill & -26.8810581$^\circ$ & 1\\
\\
$l$\dotfill     & Ecliptic Longitude\dotfill & 275.4691$^\circ$ & 1\\
$b$\dotfill     & Ecliptic Latitude\dotfill & -26.8807$^\circ$ & 1\\
\\
%$NUV$\dotfill           & GALEX $NUV$ mag.\dotfill & 13.804 $\pm$ 0.004 & 2 \\
%$FUV$\dotfill           & GALEX $FUV$ mag.\dotfill & 18.466 $\pm$ 0.056 & 2 \\
\\
%B$_T$\dotfill			&Tycho B$_T$ mag.\dotfill & 8.913$^{+0.020}_{-0.016}$		& 2	\\
%V$_T$\dotfill			&Tycho V$_T$ mag.\dotfill & 8.389$^{+0.020}_{-0.012}$		& 2	\\
%& APASS Johnson $B$ mag.\dotfill	& 13.001 $\pm$	0.02& 2	\\
%$V$\dotfill		& APASS Johnson $V$ mag.\dotfill	& 11.808 $\pm$	0.02& 2	\\
%\\
${\rm G}$\dotfill     & Gaia $G$ mag.\dotfill     &12.0665$\pm$0.0005& 1\\
%${\rm T}$\dotfill     & TESS mag.\dotfill     &10.711$\pm$0.019& 4\\
%$u'$\dotfill        & Sloan $u'$ mag.\dotfill & 14.706 $\pm$ 0.006& 3\\
%$g'$\dotfill		& APASS Sloan $g'$ mag.\dotfill	& 12.407 $\pm$ 0.02	&  2	\\
%$r'$\dotfill		& APASS Sloan $r'$ mag.\dotfill	& 11.311 $\pm$ 0.02	&  2	\\
%$i'$\dotfill		& APASS Sloan $i'$ mag.\dotfill	& 10.927 $\pm$ 0.04 &  2	\\
\\
J\dotfill			& 2MASS J mag.\dotfill & 9.469 $\pm$ 0.023	& 2	\\
H\dotfill			& 2MASS H mag.\dotfill & 8.893 $\pm$ 0.027	    &  2	\\
K$_S$\dotfill			& 2MASS ${\rm K_S}$ mag.\dotfill & 8.634 $\pm$ 0.023&  2	\\
\\
\textit{WISE1}\dotfill		& \textit{WISE1} mag.\dotfill & 8.511 $\pm$ 0.023 & 3	\\
\textit{WISE2}\dotfill		& \textit{WISE2} mag.\dotfill & 8.387 $\pm$ 0.02\ &  3 \\
\textit{WISE3}\dotfill		& \textit{WISE3} mag.\dotfill &  8.289 $\pm$ 0.016& 3	\\
\textit{WISE4}\dotfill		& \textit{WISE4} mag.\dotfill & 8.267 $\pm$ 0.083 &  3	\\
\\
$\mu_{\alpha}$\dotfill		& Gaia DR2 proper motion\dotfill & -102.75 $\pm$ 0.05 & 1 \\
                    & \hspace{3pt} in RA (mas yr$^{-1}$)	&  \\
$\mu_{\delta}$\dotfill		& Gaia DR2 proper motion\dotfill 	&  161.80 $\pm$ 0.06 &  1 \\
                    & \hspace{3pt} in DEC (mas yr$^{-1}$) &  \\
%$v\sin{i_\star}$\dotfill &  Rotational velocity (\kms) \hspace{9pt}\dotfill &  5.1 $\pm$ 0.5 & \S\ref{sec:TRES}\\
%$\feh$\dotfill &   Metallicity \hspace{9pt}\dotfill & 0.14 $\pm$ 0.08 & \S\ref{sec:TRES} \\
%$\teff$\dotfill &  Effective Temperature (K) \hspace{9pt}\dotfill &  5640 $\pm$ 50  &  \S\ref{sec:TRES}  \\
%$\log{g_{\star}}$\dotfill &  Surface Gravity (cgs)\hspace{9pt}\dotfill &  3.97 $\pm$ 0.1  &  \S\ref{sec:TRES} \\
$\pi^\ddagger$\dotfill & Gaia Parallax (mas) \dotfill & 32.18 $\pm$  0.0392$^{\dagger}$ &  1 \\
%$RV$\dotfill & Systemic radial \hspace{9pt}\dotfill  & $-5.749\pm0.065$  & \S\ref{sec:TRES} \\
%     & \hspace{3pt} velocity (\kms)  & \\
$d$\dotfill & Distance (pc)\dotfill & 31.075$\pm0.038^{\ddagger}$ & 1 \\
Spec. Type\dotfill & Spectral Type\dotfill & 	M2V$\pm$1 & 4 \\
%$A_V$\dotfill & Visual extinction (mag) & $0.015\substack{-0.015 \\+0.020}$ & \S\ref{sec:TRES} \\
%$U^{*}$\dotfill & Space Velocity (\kms)\dotfill & $-11.00 \pm 0.11$  & \S\ref{sec:uvw} \\
%$V$\dotfill       & Space Velocity (\kms)\dotfill & $-9.36 \pm 0.10$ & \S\ref{sec:uvw} \\
%$W$\dotfill       & Space Velocity (\kms)\dotfill & $0.11\pm 0.05$ & \S\ref{sec:uvw} \\
\hline
\end{tabular}
\begin{flushleft}
 \footnotesize{ \textbf{\textsc{NOTES:}}
 The uncertainties of the photometry have a systematic error floor applied. \\
 $\dagger$ RA and Dec are in epoch J2000. The coordinates come from Vizier where the Gaia RA and Dec have been precessed to J2000 from epoch J2015.5.\\
 $\ddagger$ Values have been corrected for the -0.82 $\mu$as offset as advocated by \citet{Stassun:2018}.\\
% $*$ $U$ is in the direction of the Galactic center. \\
 References are: $^1$\citet{Gaia:2018},$^2$\citet{Cutri:2003}, $^3$\citet{Cutri:2014}, $^4$\citet{Gilbert:2020}
}
\end{flushleft}
\label{tbl:LitProps}
\end{table}

%% file: EXOFAST_Fit_new.tex
\startlongtable
\begin{table*}[!ht]
\scriptsize
\centering
\caption{Median values and 68\% confidence interval for global model of \thisstar}
\begin{tabular}{llcccc}
  \hline
  \hline
Parameter & Description (Units) & Values & & & \\
\smallskip\\\multicolumn{2}{l}{Stellar Parameters:}&\smallskip\\
~~~~$M_*$\dotfill &Mass (\msun)\dotfill &$0.415\pm0.020$\\
~~~~$R_*$\dotfill &Radius (\rsun)\dotfill &$0.424\pm0.017$\\
~~~~$L_*$\dotfill &Luminosity (\lsun)\dotfill &$0.0232^{+0.0027}_{-0.0025}$\\
~~~~$\rho_*$\dotfill &Density (cgs)\dotfill &$7.68^{+1.0}_{-0.89}$\\
~~~~$\log{g}$\dotfill &Surface gravity (cgs)\dotfill &$4.802^{+0.038}_{-0.039}$\\
~~~~$T_{\rm eff}$\dotfill &Effective Temperature (K)\dotfill &$3461\pm66$\\
~~~~$[{\rm Fe/H}]$\dotfill &Metallicity (dex)\dotfill &$-0.07\pm0.11$\\
\smallskip\\\multicolumn{2}{l}{Planetary Parameters:}&b&c&d\smallskip\\
~~~~$P$\dotfill &Period (days)\dotfill &$9.97702^{+0.00024}_{-0.00028}$&$16.051110^{+0.000062}_{-0.000063}$&$37.42475^{+0.00036}_{-0.00040}$\\
~~~~$R_P$\dotfill &Radius (\re)\dotfill &$1.037^{+0.065}_{-0.064}$&$2.65^{+0.16}_{-0.15}$&$1.144^{+0.062}_{-0.061}$\\
%~~~~$M_P$\dotfill &Mass (\me)\dotfill &$1.25^{+1.00}_{-0.35}$&$7.9^{+2.7}_{-1.8}$&$1.94^{+0.69}_{-0.57}$\\
~~~~$T_C$\dotfill &Time of conjunction (\bjdtdb)\dotfill &$2458331.3537^{+0.0059}_{-0.0032}$&$2458340.08813\pm0.00095$&$2458330.4754^{+0.0047}_{-0.0041}$\\
~~~~$T_0^\dagger$\dotfill &Optimal conjunction Time (\bjdtdb)\dotfill &$2458490.9867^{+0.0027}_{-0.0029}$&$2458548.75256\pm0.00050$&$2458742.1476^{+0.0016}_{-0.0012}$\\
~~~~$a$\dotfill &Semi-major axis (AU)\dotfill &$0.0677\pm0.0011$&$0.0929\pm0.0015$&$0.1633^{+0.0026}_{-0.0027}$\\
~~~~$i$\dotfill &Inclination (Degrees)\dotfill &$89.66^{+0.24}_{-0.29}$&$88.868^{+0.083}_{-0.10}$&$89.79^{+0.14}_{-0.12}$\\
~~~~$e$\dotfill &Eccentricity \dotfill &$0.081^{+0.095}_{-0.058}$&$0.078^{+0.075}_{-0.056}$&$0.111^{+0.14}_{-0.078}$\\
~~~~$\omega_*$\dotfill &Argument of Periastron (Degrees)\dotfill &$-75^{+94}_{-84}$&$81^{+80}_{-83}$&$0^{+140}_{-130}$\\
~~~~$T_{eq}$\dotfill &Equilibrium temperature (K)\dotfill &$417\pm12$&$356\pm10.$&$268.8^{+7.7}_{-7.6}$\\
%~~~~$\tau_{\rm circ}$\dotfill &Tidal circularization timescale (Gyr)\dotfill &$24600^{+14000}_{-4600}$&$11200^{+4500}_{-3000}$&$6800000^{+3000000}_{-1400000}$\\
%~~~~$K$\dotfill &RV semi-amplitude (m/s)\dotfill &$0.67^{+0.54}_{-0.19}$&$3.61^{+1.3}_{-0.82}$&$0.68^{+0.24}_{-0.20}$\\
%~~~~$\log{K}$\dotfill &Log of RV semi-amplitude \dotfill &$-0.17^{+0.26}_{-0.14}$&$0.56^{+0.13}_{-0.11}$&$-0.17^{+0.13}_{-0.15}$\\
~~~~$R_P/R_*$\dotfill &Radius of planet in stellar radii \dotfill &$0.0224^{+0.0010}_{-0.0011}$&$0.0573^{+0.0020}_{-0.0018}$&$0.02476^{+0.00088}_{-0.00089}$\\
~~~~$a/R_*$\dotfill &Semi-major axis in stellar radii \dotfill &$34.3^{+1.5}_{-1.4}$&$47.1^{+2.0}_{-1.9}$&$82.9^{+3.5}_{-3.3}$\\
~~~~$\delta$\dotfill &Transit depth (fraction)\dotfill &$0.000503^{+0.000048}_{-0.000046}$&$0.00329^{+0.00024}_{-0.00020}$&$0.000613^{+0.000045}_{-0.000043}$\\
~~~~$Depth$\dotfill &Flux decrement at mid transit \dotfill &$0.000503^{+0.000048}_{-0.000046}$&$0.00329^{+0.00024}_{-0.00020}$&$0.000613^{+0.000045}_{-0.000043}$\\
~~~~$\tau$\dotfill &Ingress/egress transit duration (days)\dotfill &$0.00220^{+0.00029}_{-0.00017}$&$0.0138^{+0.0026}_{-0.0023}$&$0.00367^{+0.00078}_{-0.00032}$\\
~~~~$T_{14}$\dotfill &Total transit duration (days)\dotfill &$0.0943^{+0.0056}_{-0.0049}$&$0.0589^{+0.0019}_{-0.0018}$&$0.1384^{+0.0033}_{-0.0027}$\\
~~~~$T_{FWHM}$\dotfill &FWHM transit duration (days)\dotfill &$0.0920^{+0.0055}_{-0.0048}$&$0.0451^{+0.0020}_{-0.0023}$&$0.1345^{+0.0031}_{-0.0027}$\\
~~~~$b$\dotfill &Transit Impact parameter \dotfill &$0.21^{+0.19}_{-0.15}$&$0.893^{+0.017}_{-0.021}$&$0.30^{+0.19}_{-0.20}$\\
~~~~$b_S$\dotfill &Eclipse impact parameter \dotfill &$0.20^{+0.15}_{-0.14}$&$0.951^{+0.14}_{-0.074}$&$0.30^{+0.12}_{-0.19}$\\
~~~~$\tau_S$\dotfill &Ingress/egress eclipse duration (days)\dotfill &$0.00205\pm0.00023$&$0.0141^{+0.0078}_{-0.014}$&$0.00366^{+0.00041}_{-0.00038}$\\
~~~~$T_{S,14}$\dotfill &Total eclipse duration (days)\dotfill &$0.0888^{+0.0075}_{-0.0095}$&$0.052^{+0.011}_{-0.052}$&$0.138^{+0.015}_{-0.016}$\\
~~~~$T_{S,FWHM}$\dotfill &FWHM eclipse duration (days)\dotfill &$0.0867^{+0.0074}_{-0.0093}$&$0.026^{+0.023}_{-0.026}$&$0.134^{+0.015}_{-0.016}$\\
~~~~$\delta_{S,3.6\mu m}$\dotfill &Blackbody eclipse depth at 3.6$\mu$m (ppm)\dotfill &$0.068^{+0.020}_{-0.017}$&$0.085^{+0.031}_{-0.024}$&$0.00039^{+0.00019}_{-0.00013}$\\
~~~~$\delta_{S,4.5\mu m}$\dotfill &Blackbody eclipse depth at 4.5$\mu$m (ppm)\dotfill &$0.358^{+0.085}_{-0.072}$&$0.63^{+0.18}_{-0.14}$&$0.0063^{+0.0024}_{-0.0018}$\\
%~~~~$\rho_P$\dotfill &Density (cgs)\dotfill &$5.98^{+4.4}_{-0.99}$&$2.31^{+0.78}_{-0.51}$&$6.6^{+2.9}_{-1.2}$\\
%~~~~$logg_P$\dotfill &Surface gravity \dotfill &$3.046^{+0.25}_{-0.094}$&$3.04^{+0.12}_{-0.10}$&$3.14^{+0.15}_{-0.10}$\\
%~~~~$\Theta$\dotfill &Safronov Number \dotfill &$0.0137^{+0.011}_{-0.0032}$&$0.047^{+0.015}_{-0.010}$&$0.046^{+0.017}_{-0.011}$\\
~~~~$\fave$\dotfill &Incident Flux (\fluxcgs)\dotfill &$0.00681^{+0.00083}_{-0.00077}$&$0.00362^{+0.00043}_{-0.00039}$&$0.00116^{+0.00014}_{-0.00013}$\\
~~~~$T_P$\dotfill &Time of Periastron (\bjdtdb)\dotfill &$2458327.3^{+3.1}_{-2.8}$&$2458324.0^{+3.3}_{-3.1}$&$2458297^{+10.}_{-12}$\\
~~~~$T_S$\dotfill &Time of eclipse (\bjdtdb)\dotfill &$2458326.36^{+0.49}_{-0.51}$&$2458332.07^{+0.73}_{-0.68}$&$2458311.8^{+3.3}_{-3.2}$\\
~~~~$T_A$\dotfill &Time of Ascending Node (\bjdtdb)\dotfill &$2458328.80^{+0.23}_{-0.44}$&$2458336.23^{+0.56}_{-0.34}$&$2458321.1^{+1.7}_{-2.1}$\\
~~~~$T_D$\dotfill &Time of Descending Node (\bjdtdb)\dotfill &$2458333.90^{+0.44}_{-0.24}$&$2458343.96^{+0.34}_{-0.55}$&$2458339.8^{+2.1}_{-1.7}$\\
~~~~$e\cos{\omega_*}$\dotfill & \dotfill &$-0.000^{+0.078}_{-0.080}$&$0.000^{+0.071}_{-0.067}$&$0.00\pm0.14$\\
~~~~$e\sin{\omega_*}$\dotfill & \dotfill &$-0.027^{+0.043}_{-0.085}$&$0.031^{+0.077}_{-0.044}$&$-0.001^{+0.056}_{-0.086}$\\
~~~~$M_P\sin i$\dotfill &Minimum mass (\me)\dotfill &$1.25^{+1.00}_{-0.35}$&$7.9^{+2.7}_{-1.8}$&$1.94^{+0.69}_{-0.57}$\\
~~~~$M_P/M_*$\dotfill &Mass ratio \dotfill &$0.0000091^{+0.0000073}_{-0.0000025}$&$0.000057^{+0.000020}_{-0.000013}$&$0.0000140^{+0.0000051}_{-0.0000041}$\\
~~~~$d/R_*$\dotfill &Separation at mid transit \dotfill &$35.1^{+3.5}_{-2.5}$&$45.3^{+3.6}_{-4.3}$&$82.2^{+7.9}_{-7.6}$\\
%~~~~$P_T$\dotfill &A priori non-grazing transit prob \dotfill &$0.0278^{+0.0022}_{-0.0025}$&$0.0208^{+0.0022}_{-0.0015}$&$0.0119^{+0.0012}_{-0.0010}$\\
%~~~~$P_{T,G}$\dotfill &A priori transit prob \dotfill &$0.0291^{+0.0022}_{-0.0026}$&$0.0234^{+0.0025}_{-0.0017}$&$0.0125^{+0.0013}_{-0.0011}$\\
%~~~~$P_S$\dotfill &A priori non-grazing eclipse prob \dotfill &$0.0297^{+0.0026}_{-0.0017}$&$0.0194\pm0.0012$&$0.01195^{+0.0015}_{-0.00068}$\\
%~~~~$P_{S,G}$\dotfill &A priori eclipse prob \dotfill &$0.0311^{+0.0027}_{-0.0018}$&$0.0218\pm0.0014$&$0.01256^{+0.0015}_{-0.00071}$\\
\smallskip\\\multicolumn{2}{l}{Wavelength Parameters:}&z'&$4.5\mu m$&TESS\smallskip\\
~~~~$u_{1}$\dotfill &linear limb-darkening coeff \dotfill &$0.36^{+0.36}_{-0.25}$&$0.128^{+0.19}_{-0.095}$&$0.20^{+0.12}_{-0.11}$\\
~~~~$u_{2}$\dotfill &quadratic limb-darkening coeff \dotfill &$0.02^{+0.31}_{-0.26}$&$0.08^{+0.24}_{-0.14}$&$0.48\pm0.13$\\
~~~~$A_D$\dotfill &Dilution from neighboring stars \dotfill &--&--&$-0.000\pm0.011$\\
\smallskip\\\multicolumn{2}{l}{Transit Parameters:}&TESS &Spitzer UT 20191022 (4.5$\mu$m)&LCO SAAO UT 20191101 (z') & Spitzer UT 20200105 (4.5$\mu$m) \smallskip\\
~~~~$\sigma^{2}$\dotfill &Added Variance \dotfill &$0.000000091^{+0.000000036}_{-0.000000035}$&$0.0000000050^{+0.0000000082}_{-0.0000000038}$&$0.00000266^{+0.00000042}_{-0.00000037}$&$0.0000000048^{+0.0000000079}_{-0.0000000036}$\\
~~~~$F_0$\dotfill &Baseline flux \dotfill &$1.000072^{+0.000017}_{-0.000016}$&$1.000063^{+0.000040}_{-0.000041}$&$0.99989\pm0.00017$&$1.000078^{+0.000041}_{-0.000040}$\\
\end{tabular}
 \begin{flushleft} 
  \footnotesize{ 
    \noindent\textbf{\textsc{\hspace{0.75in}NOTES:}}
See Table 3 in \citet{Eastman:2019} for the definition and explanation of the derived and fitted parameters in EXOFASTv2.\\
Equilibrium temperature is calculated assuming zero albedo and perfect heat redistribution: $T_{\rm eq} = T_{\rm eff}\sqrt{\frac{R_*}{2a}}$ \\
The derived secondary eclipse depths assume a Bond albedo of zero.\\
$^\dagger$Minimum covariance with period.\\
All values in this table for the secondary occultation of \thisstar\ b are predicted values from our global analysis. }
 \end{flushleft}
\label{tbl:exofastv2}
\end{table*}
%\end{document}